\documentclass[11pt, oneside]{article}   	

\usepackage{amsfonts}
\usepackage{amssymb}
\usepackage{amsmath}
\usepackage{graphicx}
\usepackage{bm}

\usepackage{makecell}

\usepackage[T1]{fontenc}

\usepackage{natbib}

\usepackage{mathtools}
\usepackage{geometry}                		
\usepackage{authblk}						
\usepackage{float}
\usepackage{array}
\usepackage{rotating}
\usepackage{multirow}

\usepackage{tabularx}
\usepackage{xspace}

\usepackage[font={sf}]{caption}

\PassOptionsToPackage{hyphens}{url}\usepackage[colorlinks=true,allcolors=black,breaklinks=true]{hyperref}

\newcommand{\MATLAB}{\textsc{Matlab}\xspace}

\newcommand{\Cpp}{C\texttt{++}\xspace}
\newcommand{\bs}{\mathbf{s}}
\newcommand{\bX}{\mathbf{X}}
\newcommand{\btheta}{\boldsymbol{\theta}}

\newcommand{\cD}{\mathcal{D}}
\newcommand{\cQ}{\mathcal{Q}}

\newcommand{\trans}{{^\text{\tiny T}}}

\newcolumntype{C}[1]{>{\centering\arraybackslash}m{#1}}
\providecommand{\keywords}[1]{\textit{Keywords:} #1}

\title{
Pushing the Limit: A Hybrid Parallel Implementation of the Multi-resolution Approximation for Massive Data}
\author[1]{Huang Huang \thanks{huangh@ucar.edu}}

\author[2]{Lewis R. Blake \thanks{lblake@mines.edu}}
\author[3]{Dorit Hammerling \thanks{hammerling@mines.edu}}
\affil[1]{National Center for Atmospheric Research, Boulder, CO}
\affil[2]{Colorado School of Mines, Golden, CO}
\affil[3]{Colorado School of Mines, Golden, CO}

\begin{document}
\maketitle

\begin{center}
Published in \textit{NCAR Technote} (NCAR/TN-558+STR)

\url{http://dx.doi.org/10.5065/nnt6-q689}
\vspace{1cm}
\end{center}

\begin{abstract}

The multi-resolution approximation (MRA) of Gaussian processes was recently proposed to conduct likelihood-based inference for massive spatial data sets. An advantage of the methodology is that it can be parallelized. We implemented the MRA in \Cpp  for both serial and parallel versions. In the parallel implementation, we use a hybrid parallelism that employs both distributed and shared memory computing for communications between and within nodes by using the Message Passing Interface (MPI) and OpenMP, respectively. The performance of the serial code is compared between the \Cpp and \MATLAB implementations over a small data set on a personal laptop. The \Cpp parallel program is further carefully studied under different configurations by applications to data sets from around a tenth of a million to 47 million observations. We show the practicality of this implementation by demonstrating that we can get quick inference for massive real-world data sets. The serial and parallel \Cpp code can be found at \url{https://github.com/hhuang90}.

\smallskip

\vspace{1cm}

\end{abstract}

\newpage 

\tableofcontents



\newpage

\section{Introduction and Background}

\subsection{Multi-resolution Approximation}
The multi-resolution approximation (MRA) was recently proposed by \cite{Matthias2017} to handle massive spatial data sets modeled with Gaussian processes. We present a brief summary of the methodology here before moving to our implementation details.

Let $X(\bs)$ be the random process for $\bs\in\mathcal{D}$, where $\mathcal{D}$ is the physical domain of interest, typically two-dimensional. Under the Gaussian process assumption, the random process $X(\bs)\sim\hbox{GP}\big(\mu(\bs),C(\cdot,\cdot)\big)$, where $\mu(\bs)$ and $C(\cdot,\cdot)$ are the mean function and covariance function, respectively. Without loss of generality, we can assume $\mu(\bs)$ is zero because the true mean is always estimated in advance. The covariance function  $C(\cdot,\cdot)$ is a positive definite function with parameter(s) $\btheta$.

When the random process is observed at a finite set of locations $\mathcal{S}=\{\bs_1,\bs_2,\ldots,\bs_n\}$,
the random vector $\bX(\mathcal{S})=\big(X(\bs_1),X(\bs_2),\ldots,X(\bs_n)\big)\trans$
follows a multivariate normal distribution $\bX(\mathcal{S})\sim N\big(0,\Sigma(\mathcal{S})\big)$, where the variance-covariance matrix $\Sigma(\mathcal{S})$ is given by
\[
\Sigma(\mathcal{S})=C(\mathcal{S},\mathcal{S})=\left[
\begin{array}{cccc}
C(\bs_1,\bs_1) & C(\bs_1,\bs_2) & \ldots & C(\bs_1,\bs_n) \\
C(\bs_2,\bs_1) & C(\bs_2,\bs_2) & \ldots & C(\bs_2,\bs_n) \\
\vdots & \vdots & \vdots & \vdots \\
C(\bs_n,\bs_1) & C(\bs_n,\bs_2) & \ldots & C(\bs_n,\bs_n) \\
\end{array}
\right].
\]

When $n$ is large, it is computationally prohibitive to conduct inference from the data. The MRA proposes an approximate random process and substitutes it for the original random process to make inference. The approximation is based on a multi-resolution partitioning of the physical domain. Let $\cD_1=\cD$, and then for any natural number $M>1$ for the number of total levels of the multi-resolution partitioning, each region is divided into $J$ subregions, i.e., 
\[
\cD_{1,j_2,\ldots,j_{m-1}}=\bigcup\limits^J_{j_m=1}\cD_{1,j_2,\ldots,j_m}, \quad j_2,\ldots,j_{m-1}=1,\ldots,J,\quad m=2,\ldots,M.
\]

For each region $\cD_{1,j_2,\ldots,j_m}$, we place $r$ knots that are the centers of basis functions to be discussed later in Section~\ref{subsec:buildStructure}. $\cQ_{1,j_2,\ldots,j_m}$ denotes the set of the $r$ knots in $\cD_{1,j_2,\ldots,j_m}$, and $\cQ_{[m]}$ denotes all the sets $\bigcup^J_{j_m=1}\cQ_{1,j_2,\ldots,j_m}$ at level $m$. At level $M$, $\cQ_{1,j_2,\ldots,j_M}$ is chosen to be the set of observation locations in $\cD_{1,j_2,\ldots,j_M}$.

Then the multi-resolution approximation uses the predictive process~\citep{Banerjee2008} iteratively as follows. $\tau_1(\bs)$ is the predictive process that approximates $X(\bs)$ based on $\cQ_{[1]}$ at level 1, i.e, $\tau_1(\bs)=\mathbb{E}[X(\bs)\mid X(\cQ_{[1]})]$. The remainder process is approximated at level 2 by assuming independence among $\cD_{1,j_2},j_2=1,\ldots,J$, and is denoted as $\delta_2(\bs)=\{X(\bs)-\tau_1(\bs)\}_{\langle 2\rangle}$.
The notation about the subscript $\langle m \rangle$ is that for a generic Gaussian process $Y(\bs)\sim\hbox{GP}(0,K(\cdot,\cdot))$, $Y(\bs)_{\langle m\rangle}$ is a Gaussian process such that $Y(\bs)_{\langle m \rangle}\sim\hbox{GP}(0,K_{(m)}(\cdot,\cdot))$ where $K_{(m)}(\bs_1,\bs_2)=K(\bs_1,\bs_2)$ if $\bs_1$ and $\bs_2$ are in the same region at level $m$, and $K_{(m)}(\bs_1,\bs_2)=0$ otherwise. 
$\tau_2(\bs)$ is the predictive process that approximates $\delta_2(\bs)$ based on $\cQ_{[2]}$, i.e., $\tau_2(\bs)=\mathbb{E}[\delta_2(\bs)\mid \delta_2(\cQ_{[2]})]$. Repeat this procedure to level $M$. For clarity, these processes are given below,
\[
\begin{array}{ccc}
&& \tau_1(\bs)=\mathbb{E}[X(\bs)\mid X(\cQ_{[1]})], \\
\delta_2(\bs)=\{X(\bs)-\tau_1(\bs)\}_{\langle 2\rangle} &,& \tau_2(\bs)=\mathbb{E}[\delta_2(\bs)\mid \delta_2(\cQ_{[2]})],\\
\delta_3(\bs)=\{\delta_2(\bs)-\tau_2(\bs)\}_{\langle 3\rangle} &,& \tau_3(\bs)=\mathbb{E}[\delta_3(\bs)\mid \delta_3(\cQ_{[3]})],\\
\delta_4(\bs)=\{\delta_3(\bs)-\tau_3(\bs)\}_{\langle 4\rangle} &,& \tau_4(\bs)=\mathbb{E}[\delta_4(\bs)\mid \delta_4(\cQ_{[4]})],\\
\vdots&,&\vdots\\
\delta_{M}(\bs)=\{\delta_{M-1}(\bs)-\tau_{M-1}(\bs)\}_{\langle M\rangle} &,& \tau_M(\bs)=\mathbb{E}[\delta_M(\bs)\mid \delta_M(\cQ_{[M]})].\\
\end{array}
\]
Hence,
\[
\begin{array}{rcl}
X(\bs) & \approx & \tau_1(\bs)+\delta_2(\bs) \\
& \approx &\tau_1(\bs)+\tau_2(\bs)+\delta_3(\bs)\\
&\approx &\tau_1(\bs)+\tau_2(\bs)+\ldots+\tau_{M-1}(\bs)+\delta_M(\bs)\\
&\approx &\tau_1(\bs)+\tau_2(\bs)+\ldots+\tau_{M-1}(\bs)+\tau_M(\bs).\\
\end{array}
\]
Under the assumption that $X(\bs)\sim\hbox{GP}\big(0,C(\cdot,\cdot)\big)$, it can be shown that each approximated remainder process is also a Gaussian process with $\delta_m(\bs)\sim\hbox{GP}\big(0,C_m(\cdot,\cdot)\big),m=2,\ldots,M$. In addition, the covariance functions $C_m(\cdot,\cdot)=0,m=2,\ldots,M$ for $\bs_1,\bs_2$ not in the same region at level $m$ and have the forms below for $\bs_1,\bs_2\in \cD_{1,j_2,\ldots,j_m}$,
\[
\begin{array}{rcl}
C_2(\bs_1,\bs_2) & = & C(\bs_1,\bs_2)- C(\bs_1,\cQ_{1})\trans C(\cQ_{1},\cQ_{1})^{-1}C(\bs_2,\cQ_{1}), \\
C_3(\bs_1,\bs_2) & = & C_2(\bs_1,\bs_2)- C_2(\bs_1,\cQ_{1,j_2})\trans C_2(\cQ_{1,j_2},\cQ_{1,j_2})^{-1}C_2(\bs_2,\cQ_{1,j_2}), \\
& \vdots &  \\
C_M(\bs_1,\bs_2) & = & C_{M-1}(\bs_1,\bs_2)- C_{M-1}(\bs_1,\cQ_{1,j_2,\ldots,j_{M-1}})\trans\\ &&C_{M-1}(\cQ_{1,j_2,\ldots,j_{M-1}},\cQ_{1,j_2,\ldots,j_{M-1}})^{-1}C_{M-1}(\bs_2,\cQ_{1,j_2,\ldots,j_{M-1}}).  \\
\end{array}
\]
Writing $C_1(\bs_1,\bs_2) = C(\bs_1,\bs_2)$, then the predictive process $\tau_m(\bs)$ in each level can be written as 
\[
\tau_m(\bs)=C_m(\bs,\cQ_{1,j_2,\ldots,j_m})\trans\boldsymbol{\eta}_{1,j_2,\ldots,j_m}, \hbox{~for~} \bs\in\cD_{1,j_2,\ldots,j_m},
\]
where the random vector $\boldsymbol{\eta}_{1,j_2,\ldots,j_m}\sim N\big(\boldsymbol{0},C_m(\cQ_{1,j_2,\ldots,j_m},\cQ_{1,j_2,\ldots,j_m})^{-1}\big)$. In addition, $C_m(\bs,\cQ_{1,j_2,\ldots,j_m})$ is a vector-valued function of dimension $r$ on domain $\cD_{1,j_2,\ldots,j_m}$. We call each component a basis function, and it is easy to observe that each basis function achieves its maximum value at the corresponding knot in $\cQ_{1,j_2,\ldots,j_m}$.

\subsection{Strategy of Building the Structure}\label{subsec:buildStructure}
\begin{figure}[htbp]
	\centering
	\includegraphics[width=\linewidth]{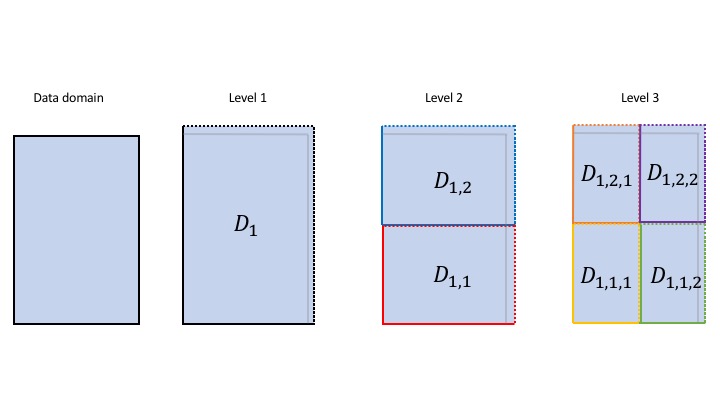}
	\caption{Illustration of building the multi-resolution structure for $J=2$, $M=3$. Dashed lines indicate open boundaries and solid lines indicate closed boundaries.}
	\label{fig:buildStructure}
\end{figure}

Starting from the data domain at level $m=1$, we recursively partition the regions into $J$ subregions. We restrict $J$ to be either 2 or 4 for simplicity. In practice, these options are sufficient because the correlation between different subregions is ignored, and if $J$ is too large, too much correlation is discarded in a single level, which is contrary to the objective. When $J=4$, every region is partitioned into four subregions with half the length for each of the two dimensions. When $J=2$, the partition is executed along the longer dimension. Figure~\ref{fig:buildStructure} gives an example of how the multi-resolution structure is built for $J=2$.
At level $m=1$, we extend the domain by pushing the top and right boundaries outward by one percent. The reason to extend the domain at level  $m=1$ slightly is that we always treat the region as a half-closed, half-open area (data on the bottom and left boundaries are included whereas data on the top and right boundaries are excluded) as indicated in Figure~\ref{fig:buildStructure}, where the points on the dashed line are excluded. By extending the domain, the original top or right boundaries as indicated by gray in Figure~\ref{fig:buildStructure} are still included, and all the observation points on those boundaries are not ignored.
Then, from level $m=2$, each region is divided into $J=2$ subregions along the longer axis. Since the length along the $y$-axis is longer at level $m=2$, we divide the region into two subregions along the $y$-axis with equal area. The same technique applies to level $m=3$ except that we divide the region along the $x$-axis, which now has a longer length in each region.
This procedure is repeated until we get to the finest resolution level $m=M$.

\begin{figure}[htbp]
	\centering
	\includegraphics[width=\linewidth]{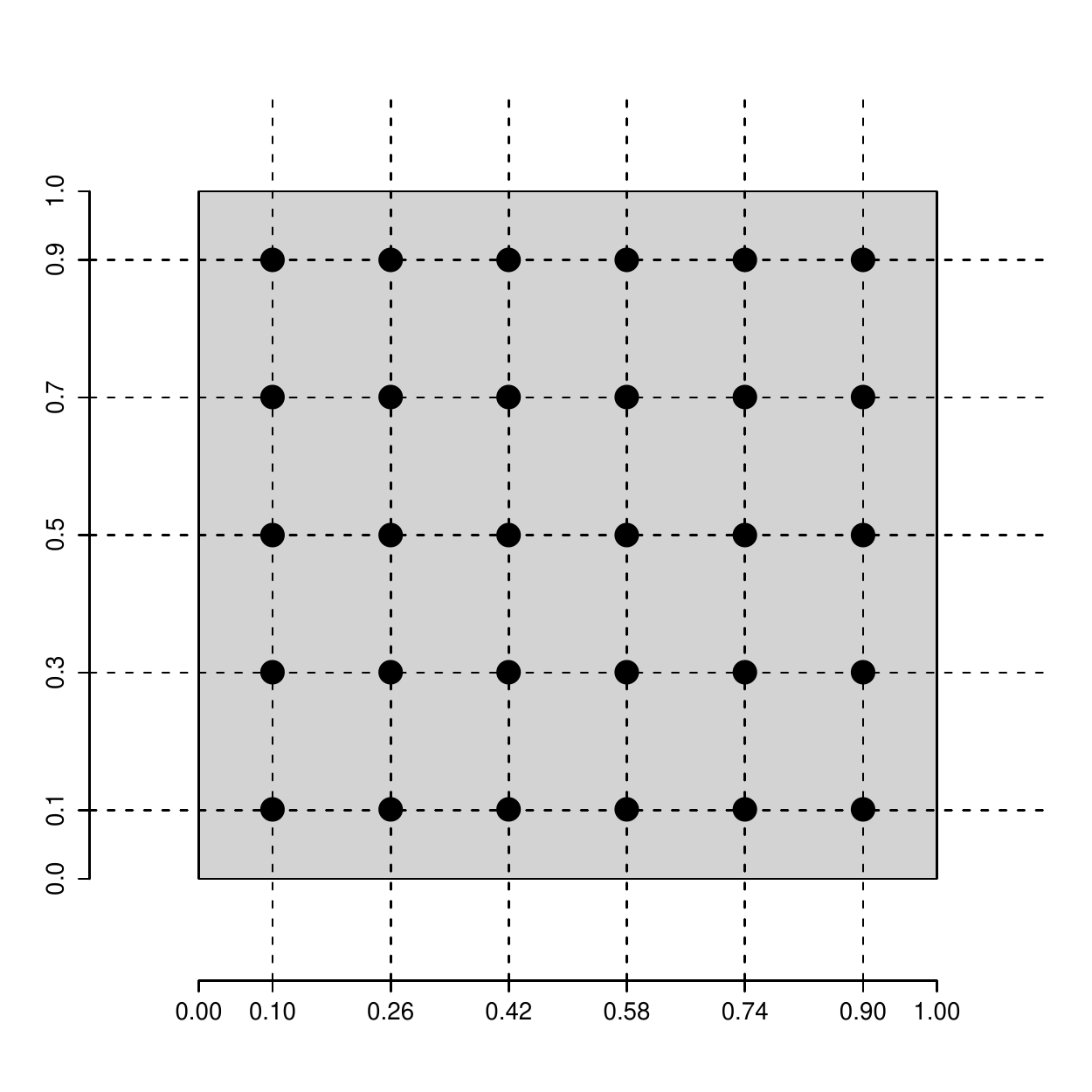}
	\caption{Illustration of the placed knots in a toy example. The region is $[0,1]\times[0,1]$ indicated in gray. Choosing $r=32$ and $\texttt{offset}=0.1$, the $\hat r=30$ placed knots are shown as solid dots. The dashed lines are the grid bars.}
	\label{fig:placeKnots}
\end{figure}

Once we have the multi-resolution structure built, we place knots, which are the centers of the basis functions, in each region. The number of knots in each region other than for the finest resolution level, i.e., levels $m=1,2,\ldots,M-1$, is specified by $r$. 
The number of knots in regions at the finest resolution level can also be chosen as $r$, resulting in a further approximation at the finest resolution level. However, in our implementation at the finest resolution level, we place the knots where the data points locate. The number of knots in each region at the finest resolution level is then automatically determined by the data set and the built structure.
Because the number of observations can be different in different regions at the finest resolution level, the number of knots can vary in regions at the finest resolution level. For levels $m=1,2,\ldots,M-1$, the knots are placed on the $\lceil\sqrt{r}\rceil\times\lfloor (r/\lceil\sqrt{r}\rceil)\rfloor$ grid, where $\lceil z\rceil$ means the minimum value that is not less than $z$, and $\lfloor z\rfloor$ means the maximum value that is not greater than $z$. The grid has equal distances between any two consecutive grid bars (see the dashed lines in Figure~\ref{fig:placeKnots}), and we use a parameter \texttt{offset} to control the smallest distance between the knots and the region boundary. If the boundary of the underlying region is $[x_{\text{min}},x_{\text{max}}]$ along $x$-dimension, then the $\lceil\sqrt{r}\rceil$ $x$-grid bars are
\[
\begin{array}{rcl}
x_1&=&x_{\text{min}}+ \texttt{offset}\times(x_{\text{max}}-x_{\text{min}}), \\
x_2&=&x_1+\dfrac{x_{\text{max}}-x_{\text{min}}-2\texttt{offset}}{\lceil\sqrt{r}\rceil-1},\\
x_3&=&x_2+\dfrac{x_{\text{max}}-x_{\text{min}}-2\texttt{offset}}{\lceil\sqrt{r}\rceil-1},\\
&\vdots&\\
x_{\lceil\sqrt{r}\rceil}&=&x_{\lceil\sqrt{r}\rceil-1}+\dfrac{x_{\text{max}}-x_{\text{min}}-2\texttt{offset}}{\lceil\sqrt{r}\rceil-1}.\\
\end{array}
\]
The locations of the $\lfloor (r/\lceil\sqrt{r}\rceil)\rfloor$ $y$-grid bars can be obtained similarly.
We see the actual number of placed knots are $\hat r=\lceil\sqrt{r}\rceil\times\lfloor (r/\lceil\sqrt{r}\rceil)\rfloor$, which is equal to $r$ when $r$ is a square number. An example is illustrated in Figure~\ref{fig:placeKnots}, where the underlying region is $[0,1]\times[0,1]$ indicated in gray. Suppose $r=32$, and then we have the $\lceil\sqrt{r}\rceil=\lceil\sqrt{32}\rceil=6$ $x$-grid bars and $\lfloor (r/\lceil\sqrt{r}\rceil)\rfloor=\lfloor (32/6)\rfloor=5$ $y$-grid bars. Therefore, we actually place $\hat r=6\times 5=30$ knots in this region. If $\texttt{offset}=0.1$, the $\hat r=30$ placed knots are shown as solid dots in Figure~\ref{fig:placeKnots}.

One restriction of placing the knots is that no knots should be exactly located at the same position. Otherwise, the variance-covariance matrix in the statistical inference will be singular. To guarantee that, one requirement is that any observation that has the same location as any knots before the finest resolution level will be eliminated. 
The chance of locations of observations coinciding with knots before the finest resolution level is low (this coincidence actually never happened in our experiments for different data sets in later sections) and we often have many observations when we use the MRA to model large data sets, so the number of observations that may need to be eliminated is negligible. The other requirement is that no knots at levels before the finest resolution level should have the same location. We use a simple strategy to prevent this collocation from happening by setting the \texttt{offset} as an irrational number, for example, $e/100$ as a default value in our code. The mathematical justification for this strategy is given in
Appendix~\ref{append:offset}.

\section{Computing Facility and Code Outline}
\subsection{Computing Facility\label{subsec:computingFacility}}
We use two kinds of computing facilities to run the code and produce the reported results. One is a personal laptop---a MacBook Pro (13-inch, early 2015) with a 3.1-GHz Intel Core i7 processor and 16 GB DDR3 memory. The other one is the high-performance computing facility including Cheyenne and Casper, built for the National Center for Atmospheric Research (NCAR) by Silicon Graphics, Inc.~(SGI). There are two types of nodes on Cheyenne---some have 45 GB usable memory and others have 109 GB usable memory. Each node has two sockets, and each socket is equipped with 18-core 2.3-GHz Intel Xeon E5-2697V4 (Broadwell) processors. 
There are multiple types of nodes on Casper as well, and we use the 22 available nodes that have 384 GB memory and two sockets equipped with 18-core 2.3-GHz Intel Xeon Gold 6140 (Skylake) processors. More information can be found at \url{https://www2.cisl.ucar.edu/resources/computational-systems/cheyenne} for Cheyenne and 
\url{https://www2.cisl.ucar.edu/resources/computational-systems/casper} for Casper.

\subsection{Code Outline}
We compare two MRA implementations---\MATLAB and \Cpp. This article focuses on introducing the \Cpp implementation, and the details of the full-layer \MATLAB implementation can be found in \cite{Blake2018}.

In the \Cpp implementation, there are two versions for the serial and parallel execution. There is little difference in the user interfaces. The only extra requirement for the parallel version is that the code should be compiled and run with Message Passing Interface (MPI) compilers and commands. Examples of MPI packages include Intel MPI, Open MPI, and SGI MPT. The users need to ensure at least one MPI package is installed before compiling and running the parallel code. In our reported results hereinafter for the parallel execution, we use the SGI MPT MPI package \texttt{mpt-2.19}\footnote{The setting \texttt{MPI\_IB\_CONGESTED} that helps alleviate network congestion but may lead to more communication overhead is disabled in our experiments.  } on the Cheyenne nodes and Intel MPI package \texttt{impi-2017.1.132} on the Casper nodes.

\subsubsection{Compiling}
The file \texttt{Makefile} is provided for both the serial and parallel versions. The users need to modify the value of ``CXX'' for the available compiler for \Cpp (the parallel code requires a \Cpp compiler that supports MPI)  before running \texttt{make} to compile. All the results in this article are from programs compiled with \texttt{clang++} on the laptop, \texttt{icpc} on Cheyenne for the serial version, and \texttt{mpicxx} as a front end of \texttt{icpc} on Cheyenne and Casper for the parallel version. In our reported results hereinafter, \texttt{clang++-x86\_64-apple-darwin17.7.0} is used on the laptop, and \texttt{icpc-17.0.1} is used on the Cheyenne and Casper nodes.

\subsubsection{Prerequisite Libraries}

There are three dependent libraries needed for our implementation---\texttt{mkl}, \texttt{armadillo}, and \texttt{dlib}. We use \texttt{mkl} as the linear algebra kernel, \texttt{armadillo} for easier manipulation of matrices~\citep{Sanderson2016}, and \texttt{dlib} for solving the optimization problems. More information about the libraries can be found at \url{https://software.intel.com/en-us/mkl}, \url{http://arma.sourceforge.net}, and \url{http://www.dlib.net}. In our reported results hereinafter, \texttt{armadillo-9.100.5} and \texttt{dlib-19.16} are used on Cheyenne nodes, Casper nodes, and the laptop. The \texttt{mkl} library \texttt{mkl-2017.0.1} is used on Cheyenne nodes and Casper nodes, and \texttt{mkl-2019.0.0} is used on the laptop.

\subsubsection{User Parameters}\label{subsubsec:userParameters}
The file \texttt{user\_parameters} is provided along with the code. This is the place for users to specify the settings for the program. The program will read parameters from this file at runtime. This file is not part of the source code, which means the code does not need to be compiled again if the parameter values in \texttt{user\_parameters} are changed.
Tables~\ref{tab:userParameters} and \ref{tab:userParameters-continued} show the user parameter names and the corresponding description. The usage of some parameters will be discussed more in Section~\ref{sec:parralel} when related parallel implementation concepts are introduced afterward.
\begin{table}[H]
	\footnotesize
	\vspace{-3mm}
	\centering
	\caption{Names and the corresponding description of user parameters.}
	\begin{tabular}{c|c}
		\hline
		\textbf{Parameter} & 
		\multicolumn{1}{|c}{
			\textbf{Description}}
		\\\hline\hline
		DATA\_FILE\_NAME & 
		\multicolumn{1}{|m{100mm}}{
			The name for the data file to be loaded. It should be a binary file. The first number is the number of observations $n$ stored as an unsigned 64-bit integer. Following are $n$ longitudes, $n$ latitudes, and $n$ observation values, each element of which is stored as a 64-bit real number.
		}
		\\\hline
		\multicolumn{1}{>{\centering\arraybackslash}m{40mm}|}{
			ELIMINATION\_DU-PLICATES\_FLAG}
		& 
		\multicolumn{1}{|m{100mm}}{
			The flag for whether to eliminate duplicates in the raw data. If it is guaranteed that there are no duplicates in the raw data, set the flag to ``false'' to save computation time. Must be either ``true'' or ``false''.
		}
		\\\hline
		OFFSET & 
		\multicolumn{1}{|m{100mm}}{
			The ratio of the smallest distance between any knot and the region boundary to the length of the region in each dimension. It should be a numerical value or ``default'' to use the value $e$/100.
		}
		\\\hline
		NUM\_PARTITIONS\_J& 
		\multicolumn{1}{|m{100mm}}{
			The number of subregions in each partitioning. It requires to be either 2 or 4 for simplicity of implementation.
		}
		\\\hline
		NUM\_KNOTS\_r & 
		\multicolumn{1}{|m{100mm}}{
			The number of knots in each region before the finest resolution level. The number of knots at the finest resolution level is automatically determined by the data and the built structure, and the number can be different in different regions at the finest resolution level.
		}
		\\\hline
		NUM\_LEVELS\_M & 
		\multicolumn{1}{|m{100mm}}{
			The number of levels. It should be a positive integer or ``default'' to use the number that is automatically determined from NUM\_PARTITIONS\_J and NUM\_KNOTS\_r by making the average number of observations per region at the finest resolution level similar to the number of knots in regions before the finest resolution level.
		}
		\\\hline
		\multicolumn{1}{>{\centering\arraybackslash}m{40mm}|}{
			PRINT\_DETAIL\_FLAG}
		& 
		\multicolumn{1}{|m{100mm}}{
			The flag for whether to print more details of the loaded data and the built structure. Must be either ``true'' or ``false''.
		}
		\\\hline
	\end{tabular}
	\label{tab:userParameters}
\end{table}

\begin{table}[H]
	\footnotesize
	\vspace{-3mm}
	\centering
	\caption{Names and the corresponding description of user parameters (continued).~$\ast$ stands for ALPHA, BETA, or TAU.}
	\begin{tabular}{c|c}
		\hline
		\textbf{Parameter} & 
		\multicolumn{1}{|c}{
			\textbf{Description}}
		\\\hline\hline
		\multicolumn{1}{>{\centering\arraybackslash}m{40mm}|}{
			CALCULATION\_MODE}
		& 
		\multicolumn{1}{|m{100mm}}{
			The type of calculations performed. It must be one of ``likelihood'' for evaluating likelihoods, ``prediction'' for making predictions, ``optimization'' for finding the optimal values of the sill parameter ALPHA, range parameter BETA, and nugget parameter TAU in the covariance function, ``build\_structure\_only'' for building the structure only where an output file ``structure\_information.txt'' will be generated to store the detailed information about the built structure. 
		}
		\\\hline
		\multicolumn{1}{>{\centering\arraybackslash}m{40mm}|}{
			PREDICTION\_LOCA-TION\_MODE}
		& 
		\multicolumn{1}{|m{100mm}}{
			The way of specifying the prediction locations. Must be one of `N' for locations in DATA\_FILE\_NAME that have NaN values, `D' for all the locations in DATA\_FILE\_NAME no matter whether the associated observation is a valid value or NaN, and `A' for locations specified in PREDICTION\_LOCATION\_FILE.
		}
		\\\hline
		\multicolumn{1}{>{\centering\arraybackslash}m{40mm}|}{
			PREDICTION\_LOCA-TION\_FILE}
		& 
		\multicolumn{1}{|m{100mm}}{
			The name for the prediction location file. Only used when PREDICTION\_LOCATION\_MODE=`A'. It should be a binary file. The first number is the number of observations $np$ that is stored as an unsigned 64-bit integer. Following are $np$ longitudes and $np$ latitudes, each element of which is stored as a 64-bit real number.
		}
		\\\hline
		\multicolumn{1}{>{\centering\arraybackslash}m{40mm}|}{
			DUMP\_PREDIC-TION\_RESULTS\_FLAG}
		&  
		\multicolumn{1}{|m{100mm}}{
			The flag for whether to dump prediction results to an output file specified by PREDICTION\_RESULT\_FILE\_NAME. Must be either "true" or "false".
		}
		\\\hline
		\multicolumn{1}{>{\centering\arraybackslash}m{40mm}|}{
			PREDICTION\_RE-SULTS\_FILE\_NAME}
		& 
		\multicolumn{1}{|m{100mm}}{
			The output file for storing prediction results. Only used when DUMP\_PREDICTION\_RESULTS\_FLAG=``true'' and CALCULATION\_MODE=``predict''. Multiple files will be generated to save the prediction results if more than one MPI process is used. Each file will save the prediction results in the following way: the first number is the number of predictions by this MPI process $nr$ as an unsigned 64-bit integer, and following are $nr$ longitudes, $nr$ latitudes, $nr$ prediction means, and $nr$ prediction variances, as 64-bit real numbers.
		}
		\\\hline
		SAVE\_TO\_DISK\_FLAG & 
		\multicolumn{1}{|m{100mm}}{
			The flag for whether to save temporary files to disk. Must be either ``true'' or ``false''.
		}
		\\\hline
		TMP\_DIRECTORY & 
		\multicolumn{1}{|m{100mm}}{
			The directory for saving the temporary files on disk. Only used when ``SAVE\_TO\_DISK''=``true''.
		}
		\\\hline
		\multicolumn{1}{>{\centering\arraybackslash}m{40mm}|}{
			DYNAMIC\_SCHE-DULE\_FLAG}
		& 
		\multicolumn{1}{|m{100mm}}{
			The flag for whether to use the dynamic scheduling of workload for each worker. Must be either ``true'' or ``false''.
		}
		\\\hline        
		ALPHA, BETA, TAU & 
		\multicolumn{1}{|m{100mm}}{
			The sill, range, and nugget parameters in the covariance function. They should be positive numerical values. Will be ignored if CALCULATION\_MODE is ``optimization'' where the optimal values that maximizes the likelihood will be found.
		}
		\\\hline
		MAX\_ITERATIONS & 
		\multicolumn{1}{|m{100mm}}{
			The maximum number of iterations allowed in the optimization. Only used when CALCULATION\_MODE=``optimization''.
		}
		\\\hline
		\multicolumn{1}{>{\centering\arraybackslash}m{40mm}|}{
			$\ast$\_LOWER\_BOUND, $\ast$\_UPPER\_BOUND}
		&
		\multicolumn{1}{|m{100mm}}{
			The lower bounds and upper bounds of sill, range, and nugget parameters in the covariance function in the optimization. Only used when CALCULATION\_MODE is ``optimization''.}
		\\\hline
		\multicolumn{1}{>{\centering\arraybackslash}m{40mm}|}{
			$\ast$\_INITIAL\_GUESS}
		&
		\multicolumn{1}{|m{100mm}}{
			The initial guesses of the sill, range, and nugget parameters in the covariance function in the optimization. Only used when CALCULATION\_MODE=``optimization''.}
		\\\hline
	\end{tabular}
	\label{tab:userParameters-continued}
\end{table}

\newpage

\subsubsection{Source Code}
There are two directories containing the source code---the directory \texttt{src} contains the source files and the directory \texttt{include} contains all the associated header files. Table~\ref{tab:sourceFiles} gives a brief description of all the source files in the directory \texttt{src}.

\begin{table}[h]
	\footnotesize
	\centering
	\caption{A brief description of all the source files in the directory \texttt{src}.~$\ast$ stands for any number of any characters.}
	\begin{tabular}{c|c}
		\hline
		\textbf{Source file}
		&
		\multicolumn{1}{|c}{
			\textbf{Description}}
		\\\hline\hline
		class\_data.cpp & 
		\multicolumn{1}{|m{100mm}}{
			Implements the class ``Data'' that contains all the variables related to the data and the function to read the data from the data file.
		}
		\\\hline
		class\_partition.cpp & 
		\multicolumn{1}{|m{100mm}}{
			Implements the class ``Partition''. The member variables store the details about the built multi-resolution structure. The member functions implement the functionality of building the structure. There are some private member functions that support the functionality.
		}
		\\\hline
		\multicolumn{1}{>{\centering\arraybackslash}m{40mm}|}{
			class\_approximation.cpp class\_approximation-$\ast$.cpp}
		& 
		\multicolumn{1}{|m{100mm}}{
			Implements the class ``Approximation''. This is the place where the MRA is executed. The member variables are about the statistical quantities in the MRA approximation. The member functions implement the functionality of creating the prior, conducting the posterior inference, and making predictions. Various private member functions support the functionality.
		}
		\\\hline
		constants.cpp & 
		\multicolumn{1}{|m{100mm}}{
			Defines all the constants or parameters, which include all the user parameters and constants throughout the program.
		}
		\\\hline
		evaluate\_covariance.cpp & 
		\multicolumn{1}{|m{100mm}}{
			Implements the covariance function. Currently the exponential covariance function is implemented. Users can modify this source file if another covariance function is preferred.
		}
		\\\hline
		optimization.cpp & 
		\multicolumn{1}{|m{100mm}}{
			Implements the function for finding the optimal values of the sill, range, and nugget parameters in the covariance function.
		}
		\\\hline
		read\_user\_parameters.cpp & 
		\multicolumn{1}{|m{100mm}}{
			Reads the specified values of the user parameters from the file ``user\_parameters''.
		}
		\\\hline
		main.cpp& 
		\multicolumn{1}{|m{100mm}}{
			The main source file that executes the functionality in the following order: initializing MPI processes if it is the parallel version, reading user parameters, loading data, building the multi-resolution structure, conducting the calculation mode (one of building the structure only, evaluating likelihoods, finding the optimal covariance function parameter values that maximize the likelihood, and making predictions), and terminating MPI processes if it is the parallel version.
		}\\\hline
	\end{tabular}
	\label{tab:sourceFiles}
\end{table}

\subsubsection{Calculation Mode}
There are four calculation modes that can be chosen---``build\_structure\_only'', ``likelihood'', ``prediction'', and ``optimization''. The ``build\_structure\_only'' mode only prints out the details about the built multi-resolution structure. The ``likelihood'' and ``prediction'' modes evaluate the likelihood and make predictions for given parameters, respectively. The ``optimization'' mode iterates the likelihood evaluations to find the optimal parameter values that maximize the likelihood. There are a variety of optimization libraries and methods that can be used. In our implementation, we use the \texttt{BOBYQA} method~\citep{Powell2009} in the library \texttt{dlib} for optimization, which does not require knowing the gradient. Users can change the optimization method by modifying the file \texttt{optimization.cpp} in the directory \texttt{src} if another optimization method is preferred.

\section{Serial Implementation Performance Comparison between \Cpp and {\MATLAB}}

\subsection{Small MODIS Data}

\subsubsection{Data Description}\label{subsubsec:smallMODIS}

\begin{figure}[ht!]
	\centering
	\includegraphics[width=\linewidth]{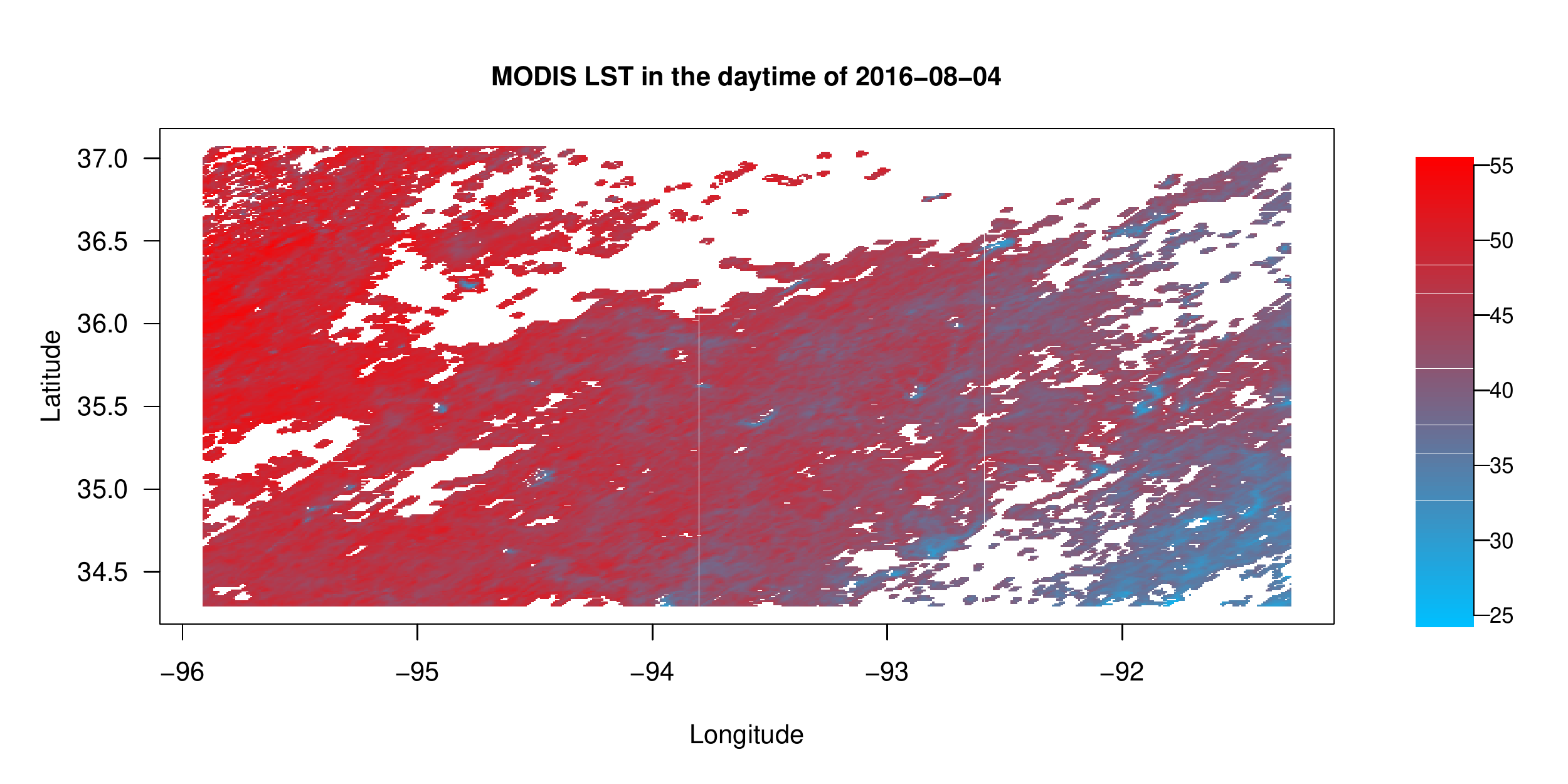}
	\caption{Small MODIS data set from \cite{Heaton2018}. MODIS land surface temperatures (LST) in Celsius in the daytime of August 4, 2016 at 105,569 locations. Note that there are some mistakes in the magnitude of the temperatures and potentially there is an accidental upward shift in the data values.}
	\label{fig:smallSatellite}
\end{figure}

In this section, we use a small satellite data set that was originally analyzed by \cite{Heaton2018}. The data set is visualized in Figure~\ref{fig:smallSatellite} showing daytime land surface temperatures on August 4, 2016, as measured by the Moderate Resolution Imaging Spectroradiometer (MODIS) sensor aboard the Terra Satellite. There are some missing values in this region on August 6, 2016, and this missing value pattern is applied to this data set on August 4, 2016, resulting in $105,569$ observations out of the $500\times300$ data locations. We use the serial code to calculate the likelihood from the $105,569$ observations and make predictions at all the $150,000$ locations. We found that there are some mistakes in the magnitude of the temperatures and potentially there is an accidental upward shift in the data values. Since the focus of this article is to show how we apply our MRA implementations, but not to draw geophysical conclusions about this specific data, we still use the original data from \cite{Heaton2018}.

\subsubsection{Results}
We consider an experiment with the setting shown in Table~\ref{tab:setup-exp1}. $M$ is the number of levels, $r$ is the number of knots specified in the regions before the finest resolution level, and $\hat r$ is the actual number of placed knots in the regions before the finest resolution level. The slight difference between $r$ and $\hat r$ comes from our partitioning strategy as described in Section~\ref{subsec:buildStructure}. We always choose the natural number that is closest to and not less than the square root of $r$ as the number of knots in one dimension. Then, the number of knots in the other dimension is determined accordingly. If $r$ is not a square number, there may be a slight difference between the specified value and the actual value $\hat r$ that is used in the program.

\begin{table}[!t]
	\centering
	\caption{Experiments setup for different values of $r$. $\hat r$ is the actual number of knots placed in each region before the finest resolution level.}
	\begin{tabular}{c||c|c|c|c|c|c|c|c|c}
		$M$& 9&	10&	11&	12&	13&	14&	15 & 16 &17\\ \hline
		$r$& 512& 256& 128& 64& 32& 16& 8 & 4 &2 \\\hline
		$\hat r$ &506& 256 & 120 & 64 & 30 & 16& 6 &4 &2
	\end{tabular}
	\label{tab:setup-exp1}
\end{table}

The experiment is executed on the laptop and one Cheyenne node with 45 GB usable memory specified in Section~\ref{subsec:computingFacility}. The running time throughout this article is measured by the elapsed real time (or the so-called wall time). 
Note that the ``optimization'' calculation mode is just a series of likelihood evaluations, so we only report the running time of evaluating likelihoods and making predictions using \Cpp and \MATLAB.
We run the serial program on the laptop only once for each of the case in Table~\ref{tab:setup-exp1}. It is noteworthy that the \MATLAB running time heavily depends on whether the operating system is just rebooted or has been used for a long time so that there is less contiguous memory space. We reboot the operating system each time before running the \MATLAB experiments on the laptop. 

\begin{figure}[t]
	\centering
	\includegraphics[width=\linewidth]{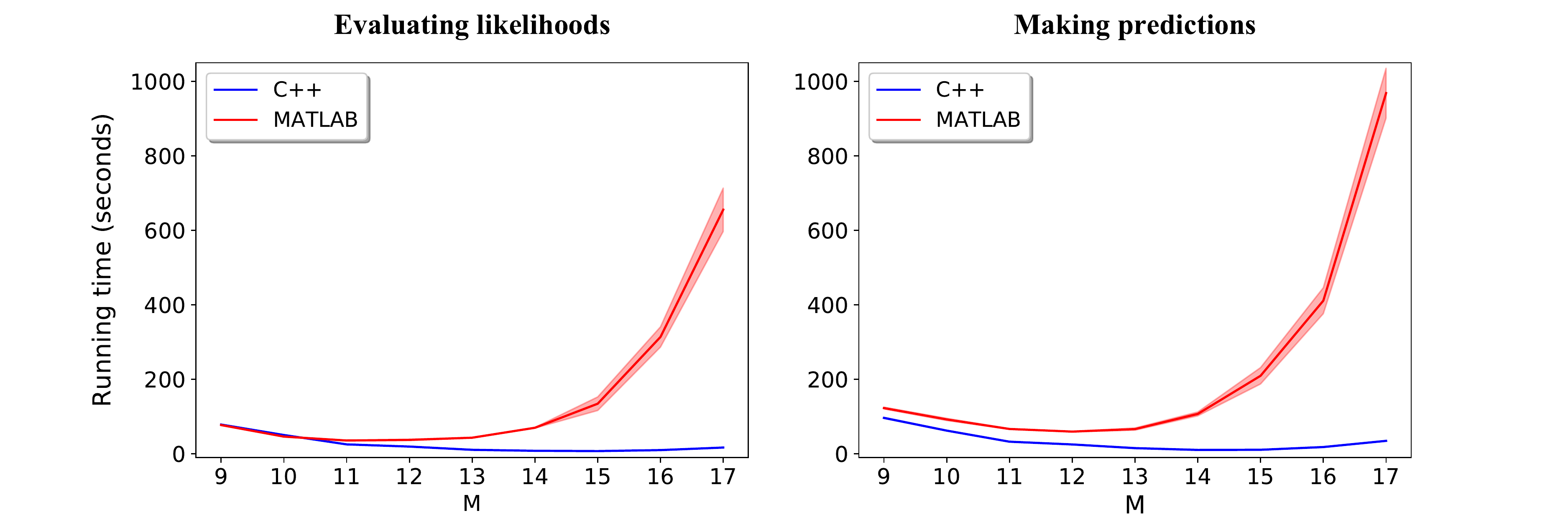}
	\caption{Running time of evaluating likelihoods and making predictions on one Cheyenne node with 45 GB usable memory using \Cpp and \MATLAB for different number of levels specified. The solid lines are the mean running time, shaded with one standard deviation above and below in the five experiments.}
	\label{fig:CheyenneSerialTime}
\end{figure}

The running time results of \Cpp and \MATLAB on the laptop are shown in Table~\ref{tab:laptop-results}. We also run the experiments for the serial program on one Cheyenne node with 45 GB usable memory. We run five experiments for each scenario in Table~\ref{tab:setup-exp1}, and the running time results on the Cheyenne node are shown in Figure~\ref{fig:CheyenneSerialTime}.
We see that the differences between the running time of \Cpp and \MATLAB are small or negligible when $M$ is small, but \Cpp outperforms \MATLAB with more significant improvement as $M$ increases.

Comparing the running time for each program over different scenarios, we see that the running time always decreases as $M$ increases at the beginning, and increases again after a certain point. A possible reason is that at the beginning, an increasing $M$ indicates more approximation with the setting in Table~\ref{tab:setup-exp1} where the sum of the number of knots over all the regions at the second finest resolution level is restricted to be the same. When $M$ reaches a point where the benefit of further approximation in terms of computational cost is negligible, the cost of more levels to be calculated from a larger $M$ dominates, resulting in the running time increasing if $M$ continues to increase, and the extra cost is more severe in the \MATLAB implementation. Note that making predictions always requires more time than evaluating likelihoods because there are more quantities to calculate.

\begin{table}[t]
	\centering
	\caption{Running time (seconds) of evaluating likelihoods and making predictions on the laptop using \Cpp and \MATLAB for different number of levels specified.}
	\begin{tabular}{c||c|c|c|c|c|c|c|c|c}
		\hline
		\multicolumn{10}{c}{\bf Evaluating likelihoods}\\\hline
		$M$& 9&	10&	11&	12&	13&	14&	15 & 16 &17\\ \hline\hline
		\Cpp & 247.97 & 87.12 & 27.59 & 15.10  & 8.43  & 7.85  & 5.81  & 8.57  & 15.37\\\hline
		\MATLAB &323.81	&155.22	&39.46&	29.92&	35.57&	54.21&	102.89&	219.01&	460.49 \\
	\end{tabular}
	
	\begin{tabular}{c||c|c|c|c|c|c|c|c|c}
		\Xhline{2\arrayrulewidth}
		\multicolumn{10}{c}{\bf Making predictions}\\\hline
		$M$& 9&	10&	11&	12&	13&	14&	15 & 16 &17\\ \hline\hline
		\Cpp& 461.95& 154.48&	46.43&	24.30&	14.02&	11.32&	9.35&	14.75&	27.77\\\hline
		\MATLAB &502.51&	233.95&	69.25&	45.94&	54.16&	81.80&	147.69&	302.34&	656.24 \\\hline
	\end{tabular}
	\label{tab:laptop-results}
\end{table}

\begin{figure}[!t]
	\centering
	\includegraphics[width=\linewidth]{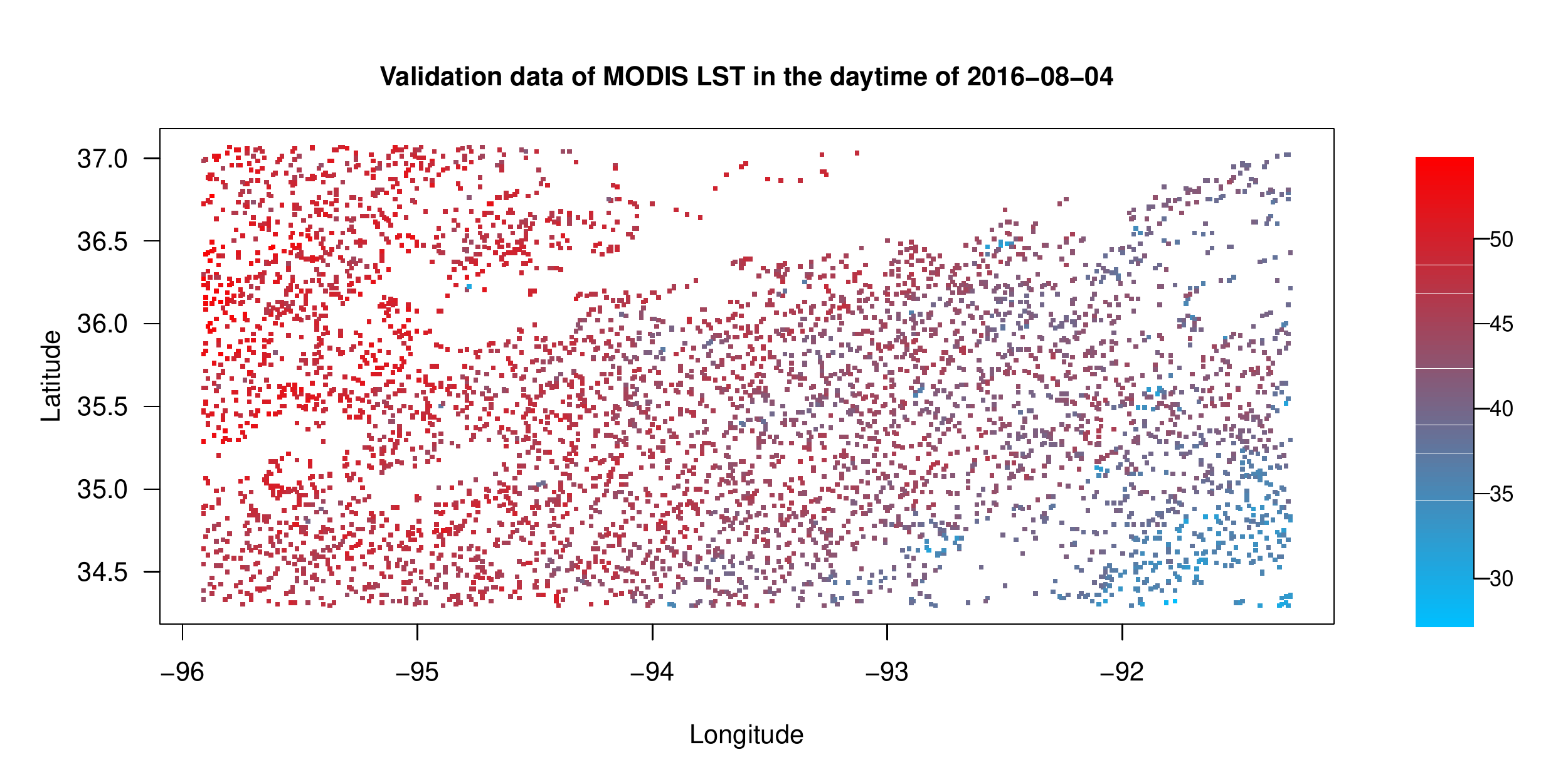}
	\caption{The randomly selected validation data points that are left out for prediction.}
	\label{fig:validationData}
\end{figure}

\begin{figure}[!t]
	\centering
	\includegraphics[width=\linewidth]{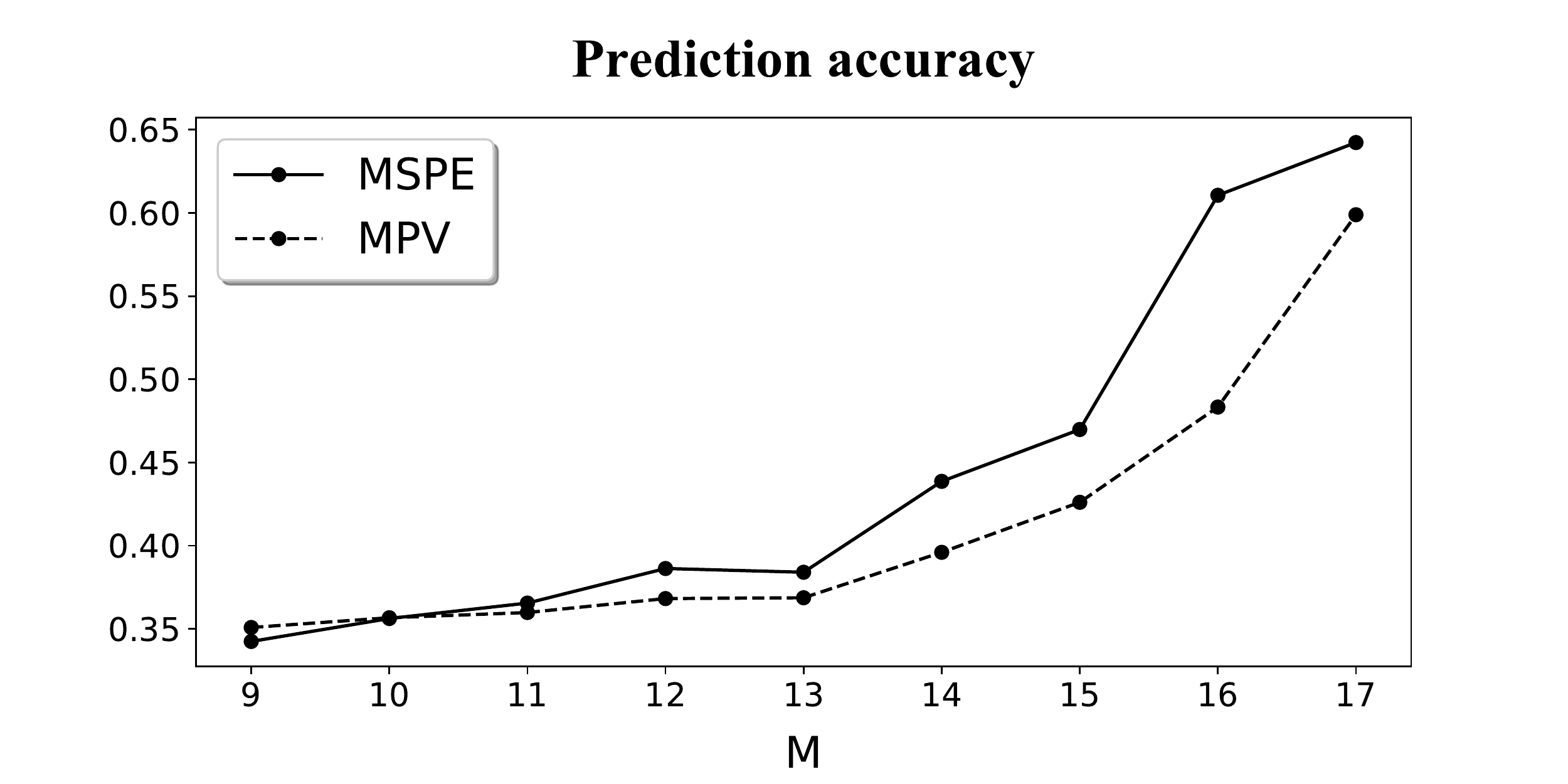}
	\caption{Mean squared prediction error (MSPE) and mean prediction variance (MPV) for different number of levels specified either using \Cpp or \MATLAB. Note that the prediction results from \Cpp and \MATLAB are exactly the same.}
	\label{fig:predictionAccuray}
\end{figure}

To show the prediction accuracy, we randomly leave out 5,000 data points from the original data set and predict the land surface temperatures at these 5,000 data points without access to the true values. The locations and the values of the data points that are left out are shown in Figure~\ref{fig:validationData}, where we see a good coverage of the domain. Then, we calculate the differences between the prediction means and the true values. The prediction variances are also available in the prediction outputs for these 5,000 data points.
The prediction means and variances from \Cpp and \MATLAB are exactly the same, supporting evidence that the implementations are correct. The mean squared prediction error (MSPE) that is the mean of the 5,000 squared differences between the prediction means and the true values, and the mean prediction variance (MPV) that is the mean of the 5,000 prediction variances, are given in Figure~\ref{fig:predictionAccuray}. It is no surprise that the accuracy generally gets worse when $M$ increases since more approximations are applied.

\section{Parallel Implementation}\label{sec:parralel}
\subsection{Parallel Mechanism\label{subsec:parallelMech}}
It is assumed in the methodology of MRA that knots in different regions at the same level are independent and only correlated with knots in regions that are ancestors or descendants. This feature allows us to develop a parallel implementation that makes use of the independence structure. With a massive data set, the inference problem is almost always memory-bounded since the memory of a node is fixed and limited. 
Almost all high-performance computing systems nowadays have an architecture such that cores in a node have direct access to the Random-access Memory (RAM) device inside the node, while communication is required if the cores try to access data in another node's RAM.
We first focus on this type of distributed architecture, allocating local memory to each core and using the Message Passing Interface (MPI) to implement the communication between cores. 

Suppose we have $p$ MPI processes and $q=J^{M-1}$ regions at the finest resolution level. Each MPI process will be assigned $\lfloor q/p\rfloor$ regions (some are assigned $\lfloor q/p\rfloor+1$ regions to ensure that every region at the finest resolution level is assigned to an MPI process). Then, all the regions that are ancestors of the assigned regions at the finest resolution level are also assigned to the same MPI process. Therefore, each MPI process only needs to allocate memory for variables and conduct calculations related to the assigned regions. Each MPI process can work independently during the procedure calculating the prior quantities and no synchronization is needed. 

However, when looping over a region where all the children are not only handled by one MPI process during posterior inference, synchronization is needed. The MPI process with the smallest rank that handles some children of this region is chosen as the ``supervisor'' of this region, and the rest as the ``worker(s)'' of the region. The ``worker(s)''  send information about their children that are not handled by the ``supervisor'' to the ``supervisor''. Then the ``supervisor'' can use information of all the children of the underlying region to calculate the relevant quantities for that region. We call a region a working region for an MPI process when the MPI process needs to calculate the associated statistical quantities of knots in that region. After synchronization, the underlying region will not be considered a working region for the ``worker(s)'' anymore. Subsequently, every ancestor that was assigned as a working region only because it is an ancestor of this underlying region and not an ancestor of other working regions will be eliminated from the working region list for the ``worker(s)'' as well.

\begin{figure}[htbp]
	\centering
	\includegraphics[width=.9\textwidth]{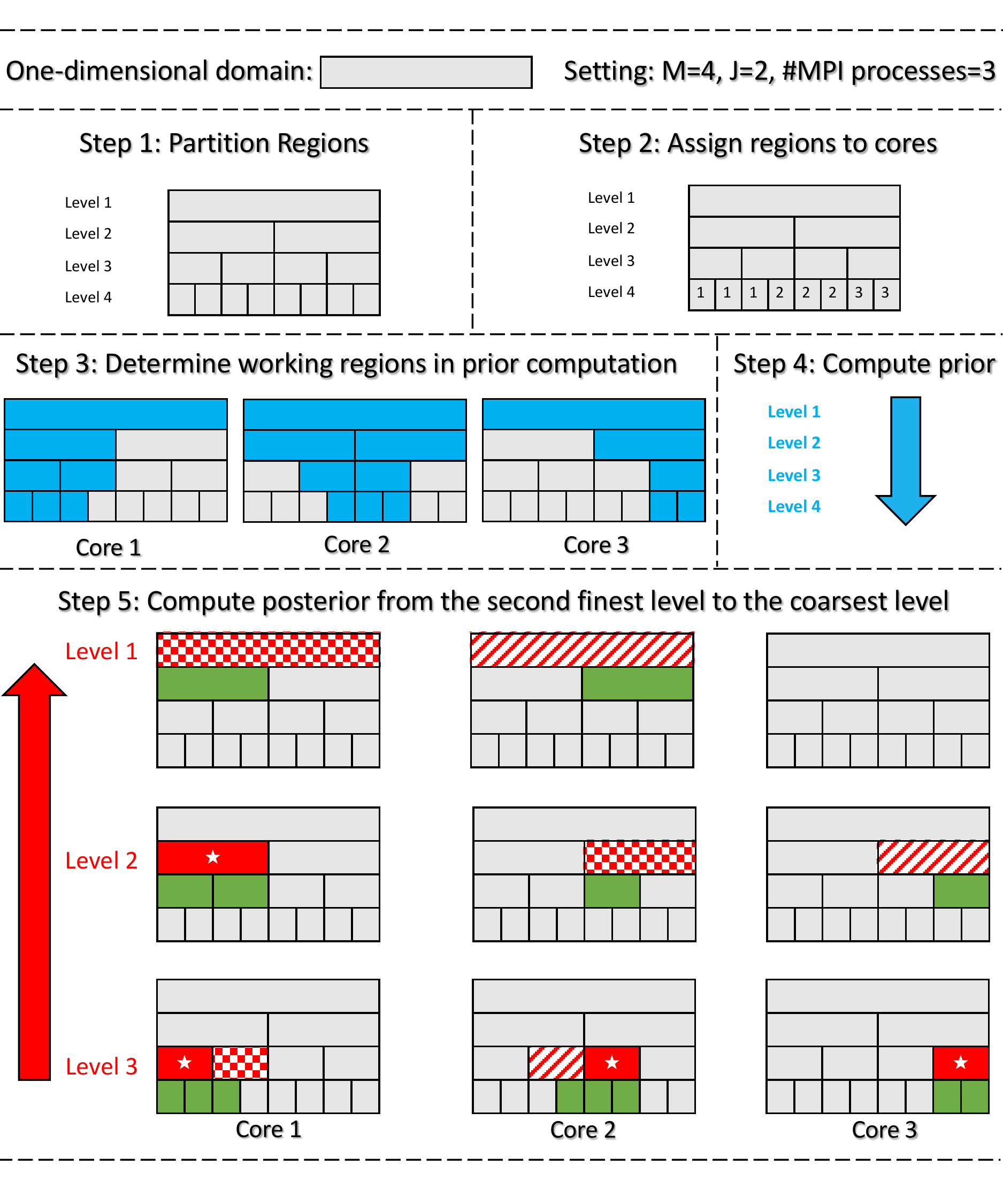}
	\caption{Illustration of the parallel MRA implementation. Solid blue indicates working regions in the prior computation. Solid green indicates working regions at one level below the current level in posterior inference. Solid red with a star indicates regions at the current level where all children were handled by this MPI process and no synchronization is needed. Checkered red and slashed red indicate regions at the current level that require synchronization, where checkered red represents the ``supervisor'' and slashed red the ``worker(s)''.}
	\label{fig:parallelProcedure}
\end{figure}

Figure~\ref{fig:parallelProcedure} illustrates the parallel mechanism on a toy example. Assume we have a one-dimensional domain. We set the number of levels $M$ to 4, the number of subregions in each partitioning $J$ to 2, and use 3 MPI processes. 

Step 1 is to build the multi-resolution structure. In each level starting from level 2, a region is divided into $J=2$ subregions. 
Step 2 is to count the number of regions each MPI process should work on and assign each MPI process their working regions. In the toy example, we have 8 regions at the finest resolution level and 3 MPI processes that can be used. This indicates each MPI process should work on at least $\lfloor 8/3 \rfloor=2$ regions. To ensure every region is assigned, the first two MPI processes are assigned one additional region, i.e., 3 regions are assigned to MPI process 1 and MPI process 2.
Step 3 is to determine all the working regions when computing prior quantities. Since there exist correlations between the assigned regions at the finest resolution level and all their ancestors, all the ancestors will be assigned as working regions as well. All the working regions are indicated in solid blue for each MPI process in Figure~\ref{fig:parallelProcedure}. 
Step 4 is to compute all the prior quantities in the order from level 1 to level 4 for a given MPI process.
Step 5 is to compute the posterior inference where the parallelism is much more complicated than in the prior computation. It starts from the second finest resolution level, which is level 3 in our example, to the coarsest resolution level, which is level 1. In Figure~\ref{fig:parallelProcedure}, solid green indicates working regions at one level below the current level. For example, for MPI process 1 at level 3, the first three regions at level 4 are shown in solid green. 
All the ancestors of green regions will be the working regions at the current level. However, there are three types of working regions at the current level.
The solid red with a star indicates all the children of this region were handled by this MPI process and no synchronization is needed. If not all the children of this region were handled by this MPI process, synchronization is required and the other two types of regions come into play. If the MPI process is the ``supervisor'' for this region, the region will be indicated in checkered red, and if the MPI process is one of the ``worker(s)'' for this region, the region will be indicated in slashed red. The ``worker(s)'' will send information related to the slashed red region to the ``supervisor''. The ``supervisor'' will gather all the information about the checkered red region from the ``worker(s)'' and compute the posterior quantities related to the checkered red region. After the synchronization, the checkered red region will still be considered a working region, which means it becomes green when the program goes one level up. However, the slashed red region will be eliminated from the working region list since all the useful information has been aggregated by the ``supervisor'' in the checkered red region, and the slashed red region will not become green when the program goes one level up.

As will be seen later in \ref{subsec:modelComplexity}, the problem is almost always memory-bounded for an extremely large data set, which means we can only use a small number of cores for MPI processes in a node due to the memory constraint. Therefore, the majority of cores in the node are left idle. To utilize them, we use the shared-memory multiprocessing mechanism to further parallelize the code by using the extra idle cores in the code wherever it is applicable. We use OpenMP to implement the shared-memory multiprocessing mechanism in our code. Then in the hybrid parallel program, MPI and OpenMP are used simultaneously for distributed-memory and shared-memory parallelism, respectively.

\subsection{Memory Requirement\label{subsec:modelComplexity}}
When dealing with an extremely large data set, before executing the program, we need to choose an appropriate model setup that the computing facility can handle. Memory allocation is the key issue since in practice the physical memory of a node is not infinite but limited.

When the number of observations $n$ and the number of levels $M$ are both large, the variable allocating the most memory in the \Cpp code for likelihood evaluation or optimization is \texttt{ATilde} in the class \texttt{Approximation}. Under the allocation and de-allocation mechanism of \texttt{ATilde} implemented in the code, the upper bound for its memory consumption is 
\begin{equation}\label{eq:upperBound}
	J^{M-1}\times M\times (M-1)\times r^2\times 2^{-28} \hbox{~GiB}.
\end{equation}
See Appendix~\ref{append:upperBoundATilde} for more details. 
The exact upper bound of the total memory consumption of the entire program should be something bigger than that of \texttt{ATilde}. However, the ratio between the two can vary from some value slightly greater than one to a much larger number, depending on many factors such as the values of $r$ and the distribution of the number of observations in regions at the finest resolution level. 
We do not intend to declare the value in Equation~\ref{eq:upperBound} as an estimate of the memory consumption bound, but it can be used as a reference to an initial impression of the total memory requirement. Users are required to monitor the memory usage for the specific data set and parameter choices if the memory constraint is an issue. We offer an option to write the values of \texttt{ATilde} in each region to disk and load the values from the disk when needed by setting \texttt{SAVE\_TO\_DISK\_FLAG} in the file \texttt{user\_parameters} to ``true''. However, this will significantly slow down the execution by the input/output operations. It is recommended to use a larger memory if possible rather than enabling this option. Nevertheless, this save-to-disk option can be used as a backup.

If the program is executed for making predictions, there is a chance for \texttt{BTilde} in the class \texttt{Approximation} to consume more memory than \texttt{ATilde}, depending on the number of predictions and how the prediction locations are distributed. Then, the estimate of the exact memory consumption of the entire program will be more complicated. Users can adjust the values of $M$ and $r$ accordingly to make the execution feasible for a given computing facility.

It is also noteworthy that when the program is executed in parallel, the memory consumption by each MPI process will be smaller than the total memory consumption since the code is implemented in a distributed fashion.

\subsection{Parallel Results}
Throughout the studies in Section~\ref{sec:parralel}, we will run our programs on the two kinds of Cheyenne nodes that have 45 GB and 109 GB usable memory per node, respectively, and one kind of Casper node that has 384 GB memory.

The actual running time of the parallel execution using MPI does not only depend on the program, the number of nodes, or the associated MPI processes per node, but also depends on how the MPI processes are placed within a node. If the number of MPI processes that are used is much smaller than the physical limit 36, say 2 or 4, the performance is highly related to the MPI process placement. However, the optimal configuration is very complicated and should be analyzed case by case due to many factors such as the different levels of cache, the memory architecture, the difference in communication speed between MPI processes in a socket and across sockets, and the amount of memory each MPI process will use for running the program. For simplicity, all the configurations in this section for the parallel study uniformly spread out MPI processes as much as possible over the 36 physical cores in a node.

Two types of parallel experiments are conducted---only using MPI or using both MPI and OpenMP. When we run the experiments for the hybrid parallel program that uses both MPI and OpenMP, we try to use up to all 36 cores of a node. We assign each MPI process an equal number of cores for OpenMP threads. The scheme of the MPI-OpenMP core assignment is given in Table~\ref{tab:coreAssignment}. Then, each MPI process will utilize more than one core as OpenMP threads except for the last case where only one core is available for each MPI process when using 32 MPI processes per node. We see that starting from the fourth row of Table~\ref{tab:coreAssignment}, where the number of MPI processes per node is 8, the total number of used cores is not 36 but 32 because 36 is not divisible by the specified number of MPI processes per node.

\begin{table}[htbp]
	\centering
	\caption{Core assignment for different choices of the number of MPI processes per node.}
	\begin{tabular}{c|c}
		\hline
		Number of MPI processes  & Number of cores used as OpenMP threads \\
		per node & for each MPI process
		\\\hline\hline
		1     & 36 \\
		2     & 18 \\
		4     & 9 \\
		8     & 4 \\
		16    & 2 \\
		32    & 1 \\
		\hline
	\end{tabular}%
	\label{tab:coreAssignment}%
\end{table}%

It is noteworthy that using more cores as OpenMP threads does not always improve performance. There is overhead for initializing the OpenMP threads for an OpenMP parallel code region. If there is too much dependence among the computations of OpenMP threads so that the OpenMP threads cannot do computations simultaneously, or the computation load for each thread is extremely unbalanced, the running time gained from the usage of OpenMP threads may be smaller than the loss by the overhead. We have optimized our code over several rounds of explorations to make it more favorable to use OpenMP parallelism. The experiment results in later sections will show that we do gain a lot for our MRA application by using OpenMP threads in addition to using MPI processes.

Another issue is how we divide the computational workload into different portions and assign each portion to each MPI process. Section~\ref{subsec:parallelMech} shows a predetermined static strategy that 
the regions at the finest resolution level are assigned to different MPI processes with the same number of regions (some MPI processes may have one more).
However, this may not lead to an even division because the number of observations in each region at the finest resolution level may vary if the data points are not uniformly spread over the domain. We offer another option to divide the regions at the finest resolution level by the square of the number of observations in each region, which means each MPI process handles regions at the finest resolution level with almost an equal value of the sum of these squares. We call this strategy ``dynamic scheduling'', and it is controlled by the variable \texttt{DYNAMIC\_SCHEDULE\_FLAG} in the file \texttt{user\_parameters}. When \texttt{DYNAMIC\_SCHEDULE\_FLAG} is ``false'', the static scheduling strategy elaborated in Section~\ref{subsec:parallelMech} is used; when \texttt{DYNAMIC\_SCHEDULE\_FLAG} is ``true'', the dynamic scheduling is applied. Similarly, when using OpenMP \texttt{for} loops, the number of loops is divided evenly for each thread by default. It is highly probable in our program that each thread may have a different workload if the data points are not uniformly spread over the domain. Thus, the execution will be slowed down by the slowest thread. Dynamic scheduling also applies here so that the number of loops for each thread is not determined in advance but each loop is assigned to the currently idle thread in the runtime. We also use the same variable \texttt{DYNAMIC\_SCHEDULE\_FLAG} to control the thread scheduling scheme for OpenMP.

\subsubsection{Small MODIS Data Set\label{subsubsec:satellite}}
\begin{figure}[t]
	\centering
	\includegraphics[width=\linewidth]{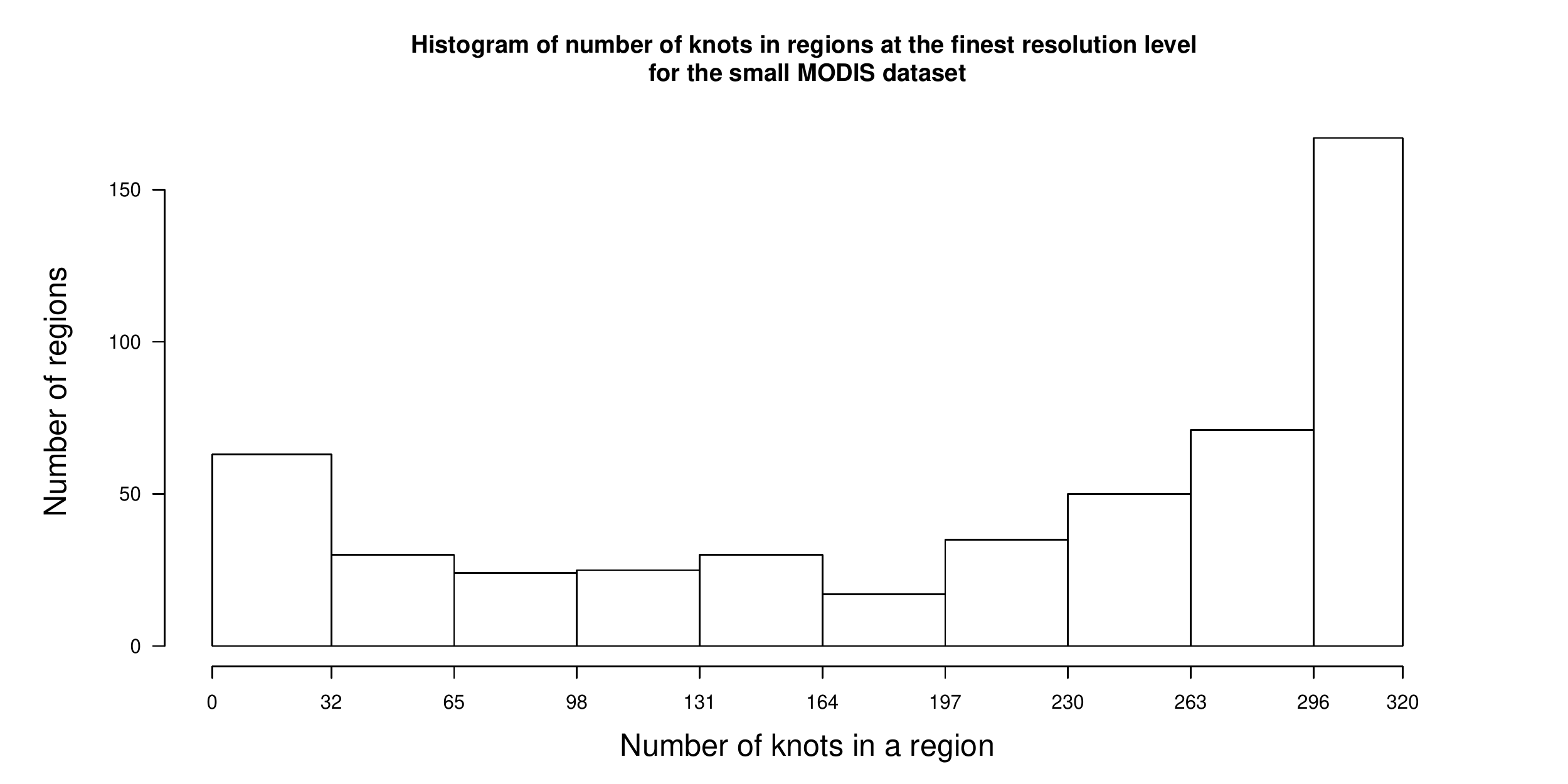}
	\caption{Histogram of the number of knots in regions at the finest resolution level for the small MODIS data set.}
	\label{fig:numberObservations-satellite}
\end{figure}

In this section, we use the same small data set described in Section~\ref{subsubsec:smallMODIS}.
We choose $M=10$ levels and $r=256$ knots in all regions before the finest resolution level to compare running time in different parallel settings. The number of partitions $J$ is set to $2$, leading to $1023$ regions in total including $512$ regions at the finest resolution level. 
In each region at the finest resolution level, the number of knots (or observations in this case) depends on how the data location is spatially distributed. Since there are $105,569$ observations and $512$ regions at the finest resolution level, we know the average number of knots is $105,569/512\approx 206$ at the finest resolution level, comparable to $r=256$.
For this data set, we found the maximum number of observations in a region at the finest resolution level is $320$, and there are 30 regions that have no observations within their designated boundaries. Figure~\ref{fig:numberObservations-satellite} shows the histogram of the number of observations in regions at the finest resolution level, where we see the data is not uniformly spread over the domain.

We run the parallel program that only uses MPI and no OpenMP threads with different total numbers of MPI processes and numbers of MPI processes per node. The experiments are run on Cheyenne nodes that have 45 GB usable memory. The results are shown in Table~\ref{tab:satellite-likelihood} for evaluating the likelihood and in Table~\ref{tab:satellite-prediction} for making predictions. The execution time for prediction is longer than for likelihood evaluation with the same setup because there are more quantities to be calculated in prediction. It is also noteworthy that the running time of making predictions would also depend on the number of locations to predict, and here we use all the $150,000$ location on the full $500\times300$ grid. In this specific case, the execution time for prediction is about $1.5$ times as long as the corresponding likelihood evaluation.

\begin{table}[tbp]
	\centering
	\caption{Mean running time (seconds) of evaluating the likelihood in the parallel implementation using MPI only with different MPI process settings for the small MODIS data set on the Cheyenne nodes with 45 GB usable memory. The scenarios along the diagonals indicated by the same color use the same number of nodes. Each experiment is run five times. Numbers in parentheses are the associated standard deviations.}
	\footnotesize
	\begin{tabular}{c|cccccc}
		\hline
		Total number  & \multicolumn{6}{c}{Number of MPI processes per node} \\
		of MPI processes & 1     & 2     & 4     & 8     & 16    & 32 \\\hline
		Serial     & \textcolor[rgb]{ 1,  0,  0}{42.96(0.19)} &       &       &       &       &  \\
		2     & \textcolor[rgb]{ 1,  .753,  0}{23.46(4.12)} & \textcolor[rgb]{ 1,  0,  0}{19.52(5.21)} &       &       &       &  \\
		4     & \textcolor[rgb]{ 0,  .69,  .941}{11.67(0.08)} & \textcolor[rgb]{ 1,  .753,  0}{8.84(0.70)} & \textcolor[rgb]{ 1,  0,  0}{10.28(1.03)} &       &       &  \\
		8     & \textcolor[rgb]{ .329,  .51,  .208}{5.87(0.02)} & \textcolor[rgb]{ 0,  .69,  .941}{4.61(0.28)} & \textcolor[rgb]{ 1,  .753,  0}{5.71(0.40)} & \textcolor[rgb]{ 1,  0,  0}{7.90(1.15)} &       &  \\
		16    & \textcolor[rgb]{ .776,  .349,  .067}{3.11(0.06)} & \textcolor[rgb]{ .329,  .51,  .208}{2.60(0.02)} & \textcolor[rgb]{ 0,  .69,  .941}{2.98(0.15)} & \textcolor[rgb]{ 1,  .753,  0}{4.42(0.32)} & \textcolor[rgb]{ 1,  0,  0}{7.14(2.62)} &  \\
		32    & \textcolor[rgb]{ .439,  .188,  .627}{1.73(0.09)} & \textcolor[rgb]{ .776,  .349,  .067}{1.42(0.04)} & \textcolor[rgb]{ .329,  .51,  .208}{1.59(0.02)} & \textcolor[rgb]{ 0,  .69,  .941}{2.57(0.09)} & \textcolor[rgb]{ 1,  .753,  0}{4.03(0.06)} & \textcolor[rgb]{ 1,  0,  0}{4.76(0.11)} \\
		64    & \textcolor[rgb]{ .125,  .216,  .392}{0.96(0.06)} & \textcolor[rgb]{ .439,  .188,  .627}{0.85(0.02)} & \textcolor[rgb]{ .776,  .349,  .067}{0.89(0.04)} & \textcolor[rgb]{ .329,  .51,  .208}{1.40(0.05)} & \textcolor[rgb]{ 0,  .69,  .941}{2.41(0.07)} & \textcolor[rgb]{ 1,  .753,  0}{2.82(0.06)} \\
		\hline
	\end{tabular}
	\label{tab:satellite-likelihood}
\end{table}

Looking at a row in Table~\ref{tab:satellite-likelihood} or Table~\ref{tab:satellite-prediction}, we see that the running time always gets shorter as the number of MPI processes per node increases at the beginning and then possibly gets longer in the end.
The reason for a shorter time at the beginning of a row is that the communication in a node is always faster than the communication across nodes. However, if too many MPI processes are used in a node, they will eventually compete for the memory and shared caches. When this competition becomes dominating, it will again slow down the running time. This phenomenon is more obvious for the larger data set shown in Section~\ref{subsubsec:amsr}. 
Looking at a column, more and more MPI processes are used in total as it goes down, indicating a higher level of parallelism. 
The running time reduces around by half as the total number of MPI processes is doubled given a fixed number of MPI processes per node.
It is also noteworthy that the values along the diagonals are from the same number of nodes used. We see that it is always beneficial to use more MPI processes per node for a given number of nodes that can be used for this data set.

This is the smallest data set we have applied our MRA implementation to. More options including using OpenMP and dynamic scheduling will be explored for a much larger data set in Section~\ref{subsubsec:amsr} when the computation becomes more intensive.

\begin{table}[tbp]
	\centering
	\caption{Mean running time (seconds) of making predictions in the parallel implementation using MPI only with different MPI process settings for the small MODIS data set on the Cheyenne nodes with 45 GB usable memory. The scenarios along the diagonals indicated by the same color use the same number of nodes. Each experiment is run five times. Numbers in parentheses are the associated standard deviations.}
	\footnotesize
	\begin{tabular}{c|cccccc}
		\hline
		Total number  & \multicolumn{6}{c}{Number of MPI processes per node} \\
		of MPI processes & 1     & 2     & 4     & 8     & 16    & 32 \\\hline
		Serial    & \textcolor[rgb]{ 1,  0,  0}{56.81(0.10)} &       &       &       &       &  \\
		2     & \textcolor[rgb]{ 1,  .753,  0}{28.90(0.26)} & \textcolor[rgb]{ 1,  0,  0}{25.33(0.86)} &       &       &       &  \\
		4     & \textcolor[rgb]{ 0,  .69,  .941}{15.39(0.04)} & \textcolor[rgb]{ 1,  .753,  0}{12.71(0.18)} & \textcolor[rgb]{ 1,  0,  0}{15.65(2.32)} &       &       &  \\
		8     & \textcolor[rgb]{ .329,  .51,  .208}{7.92(0.16)} & \textcolor[rgb]{ 0,  .69,  .941}{7.00(0.37)} & \textcolor[rgb]{ 1,  .753,  0}{9.23(0.27)} & \textcolor[rgb]{ 1,  0,  0}{13.21(2.03)} &       &  \\
		16    & \textcolor[rgb]{ .776,  .349,  .067}{4.26(0.28)} & \textcolor[rgb]{ .329,  .51,  .208}{4.07(0.05)} & \textcolor[rgb]{ 0,  .69,  .941}{5.11(0.23)} & \textcolor[rgb]{ 1,  .753,  0}{7.73(0.05)} & \textcolor[rgb]{ 1,  0,  0}{11.24(1.82)} &  \\
		32    & \textcolor[rgb]{ .439,  .188,  .627}{2.25(0.11)} & \textcolor[rgb]{ .776,  .349,  .067}{2.19(0.14)} & \textcolor[rgb]{ .329,  .51,  .208}{2.57(0.02)} & \textcolor[rgb]{ 0,  .69,  .941}{4.12(0.05)} & \textcolor[rgb]{ 1,  .753,  0}{6.22(0.72)} & \textcolor[rgb]{ 1,  0,  0}{7.66(0.09)} \\
		64    & \textcolor[rgb]{ .125,  .216,  .392}{1.37(0.09)} & \textcolor[rgb]{ .439,  .188,  .627}{1.18(0.02)} & \textcolor[rgb]{ .776,  .349,  .067}{1.59(0.08)} & \textcolor[rgb]{ .329,  .51,  .208}{2.20(0.05)} & \textcolor[rgb]{ 0,  .69,  .941}{3.88(0.10)} & \textcolor[rgb]{ 1,  .753,  0}{4.47(0.19)} \\
		\hline
	\end{tabular}
	\label{tab:satellite-prediction}
\end{table}%

\subsubsection{Medium AMSR Data Set}\label{subsubsec:amsr}
\begin{figure}[htbp]
	\centering
	\includegraphics[width=\linewidth]{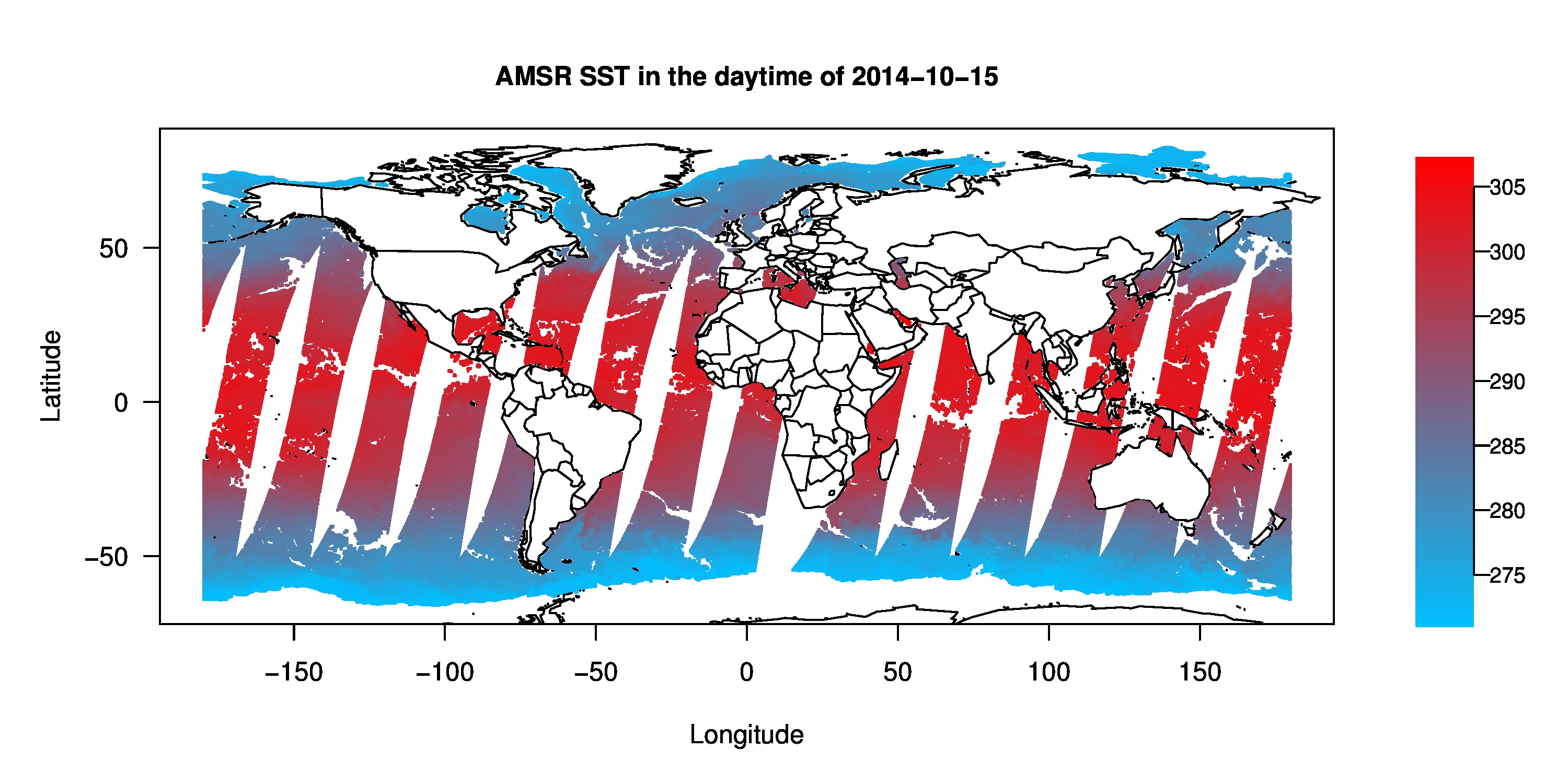}
	\caption{AMSR sea surface temperature (SST) in Kelvin in the daytime of October 15, 2014 at 2,439,827 locations.}
	\label{fig:AMSRData}
\end{figure}

There are $2,441,405$ global observations for the sea surface temperature (SST) in the daytime of October 15, 2014, from the Advanced Microwave Scanning Radiometer (AMSR) sensor aboard the Advanced Earth Observing Satellite II (ADEOS II). Some of the observations are at the same locations, and we choose only one of them, which results in $2,439,827$ unique data points. A visualization of the data set is given in Figure~\ref{fig:AMSRData}. 

Since this data set has a much larger size than that in Section~\ref{subsubsec:satellite}, we choose a larger number of levels as $M=14$ with a higher level of approximation. The number of partitions $J$ is set to $2$. Therefore, there are $16,383$ regions in total including $8,192$ regions at the finest resolution level. The number of knots is set to $r=49$ in all regions before the finest resolution level.

\begin{figure}[htbp]
	\centering
	\includegraphics[width=\linewidth]{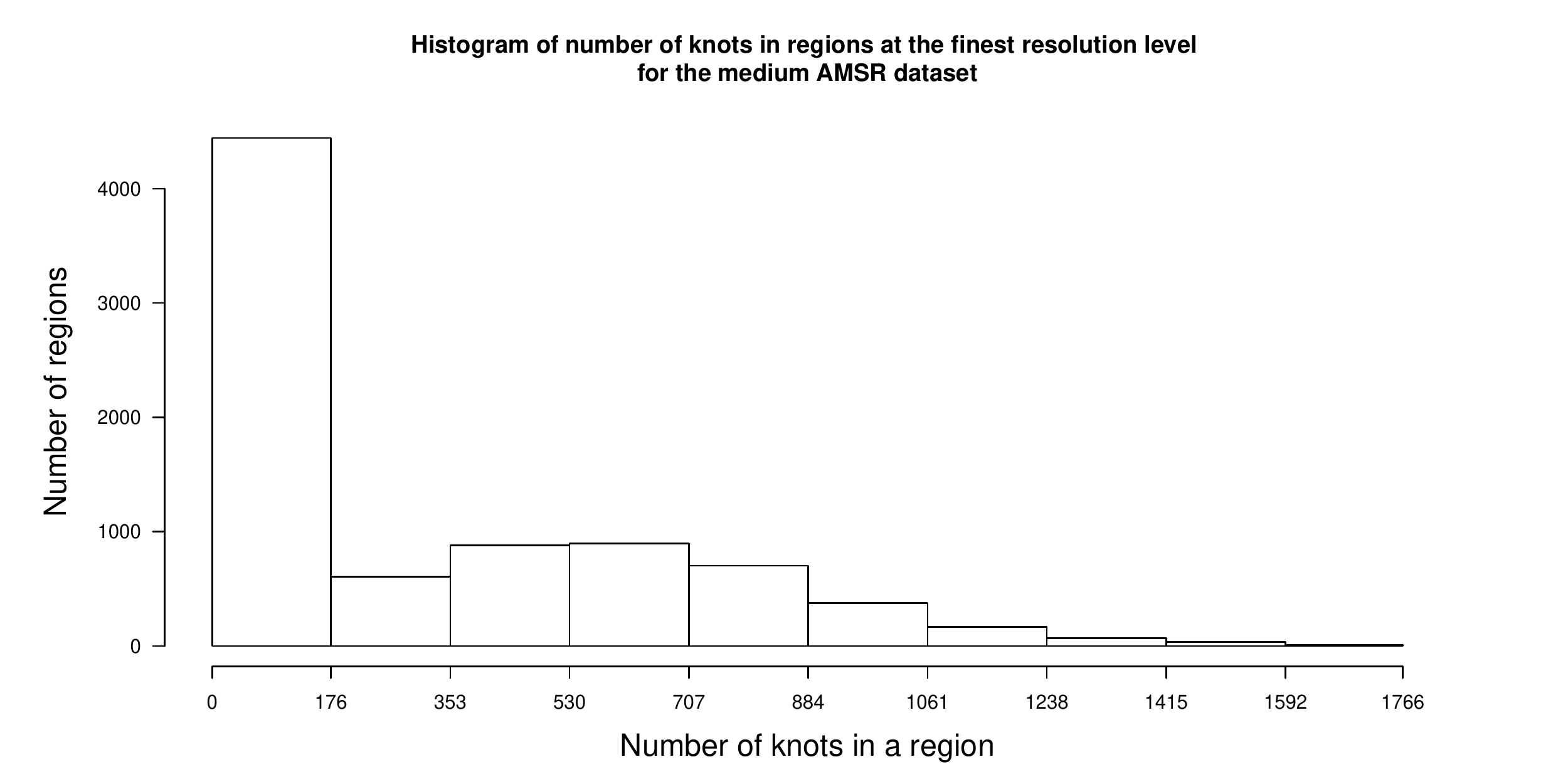}
	\caption{Histogram of the number of knots in regions at the finest resolution level for the medium AMSR data set.}
	\label{fig:numberObservations-AMSR}
\end{figure}

We know that the exact number of knots (or observations) in each region at the finest resolution level depends on the locations of the data. Since the average number of observations is $2,439,827/8,192\approx 297$ at the finest resolution level, a similar number may be used to assign $r$ for the number of knots before the finest resolution level. However, we found that this will lead to too much memory consumption, so we turned to use a moderate number for $r=49$.
The histogram of the number of observations in regions at the finest resolution level is given in Figure~\ref{fig:numberObservations-AMSR}, where we see the distribution is quite skewed in the way that a large portion of regions have smaller numbers of observations and a small portion of regions contain remarkably large numbers of observations. 

We run the parallel program that only uses MPI and no OpenMP threads on the Cheyenne nodes that have 45 GB usable memory. The results for evaluating likelihoods are shown in Table~\ref{tab:AMSR-MPI-likelihood}. The same trend happens here as in Section~\ref{subsubsec:satellite}. The running time is decreasing at the beginning and increasing in the end in a row. In a column, the running time is always decreasing when we use more MPI processes in total.
The results for making predictions at all the $2,439,827$ locations are shown in Table~\ref{tab:AMSR-MPI-prediction}. For the same setting, the ratio between the execution time of making predictions and the likelihood evaluation is greater than 2, larger than that in the small MODIS data set in Section~\ref{subsubsec:satellite} because the number of locations we chose to predict at is much larger. As stated in Section~\ref{subsec:modelComplexity}, making predictions consumes more memory than the likelihood evaluation; we see that all the settings indicated in red in Table~\ref{tab:AMSR-MPI-prediction} show insufficient memory for the execution. The reason is that all the settings in red in Table~\ref{tab:AMSR-MPI-prediction} require only one node, which only has 45 GB usable memory. However, the entire memory consumption for the predictions goes beyond that. To handle these failure cases, we use one node with 109 GB usable memory in Cheyenne to run these prediction routines, and the results are shown in Table~\ref{tab:AMSR-MPI-prediction-large-node}.

\begin{table}[htbp]
	\footnotesize
	\centering
	\caption{Mean running time (seconds) of evaluating the likelihood in the parallel implementation using MPI only with different MPI process settings for the medium AMSR data set on the Cheyenne nodes with 45 GB usable memory. The scenarios along the diagonals indicated by the same color use the same number of nodes. Each experiment is run five times. Numbers in parentheses are the associated standard deviations.}
	\begin{tabular}{c|cccccc}
		\hline
		Total number  & \multicolumn{6}{c}{Number of MPI processes per node} \\
		of MPI processes & 1     & 2     & 4     & 8     & 16    & 32 \\\hline
		Serial     & \textcolor[rgb]{ 1,  0,  0}{154.63(0.41)} &       &       &       &       &  \\
		2     & \textcolor[rgb]{ 1,  .753,  0}{83.32(0.34)} & \textcolor[rgb]{ 1,  0,  0}{66.14(0.22)} &       &       &       &  \\
		4     & \textcolor[rgb]{ 0,  .69,  .941}{45.27(0.48)} & \textcolor[rgb]{ 1,  .753,  0}{36.61(1.40)} & \textcolor[rgb]{ 1,  0,  0}{45.44(3.18)} &       &       &  \\
		8     & \textcolor[rgb]{ .329,  .51,  .208}{28.62(0.15)} & \textcolor[rgb]{ 0,  .69,  .941}{23.84(1.27)} & \textcolor[rgb]{ 1,  .753,  0}{28.37(2.03)} & \textcolor[rgb]{ 1,  0,  0}{36.97(2.69)} &       &  \\
		16    & \textcolor[rgb]{ .776,  .349,  .067}{15.63(0.23)} & \textcolor[rgb]{ .329,  .51,  .208}{13.62(1.10)} & \textcolor[rgb]{ 0,  .69,  .941}{16.85(0.26)} & \textcolor[rgb]{ 1,  .753,  0}{20.13(1.28)} & \textcolor[rgb]{ 1,  0,  0}{27.80(1.76)} &  \\
		32    & \textcolor[rgb]{ .439,  .188,  .627}{8.27(0.12)} & \textcolor[rgb]{ .776,  .349,  .067}{6.97(0.22)} & \textcolor[rgb]{ .329,  .51,  .208}{8.31(0.39)} & \textcolor[rgb]{ 0,  .69,  .941}{11.15(0.51)} & \textcolor[rgb]{ 1,  .753,  0}{14.74(0.65)} & \textcolor[rgb]{ 1,  0,  0}{21.34(0.15)} \\
		64    & \textcolor[rgb]{ .125,  .216,  .392}{4.64(0.03)} & \textcolor[rgb]{ .439,  .188,  .627}{4.39(0.14)} & \textcolor[rgb]{ .776,  .349,  .067}{5.12(0.23)} & \textcolor[rgb]{ .329,  .51,  .208}{6.75(0.36)} & \textcolor[rgb]{ 0,  .69,  .941}{8.57(0.37)} & \textcolor[rgb]{ 1,  .753,  0}{12.24(0.06)} \\
		\hline
	\end{tabular}%
	\label{tab:AMSR-MPI-likelihood}%
\end{table}%

\begin{table}[htbp]
	\centering
	\caption{Mean running time (seconds) of making predictions in the parallel implementation using MPI only with different MPI process settings for the medium AMSR data set on the Cheyenne nodes with 45 GB usable memory. The scenarios along the diagonals indicated by the same color use the same number of nodes. Each experiment is run five times. Numbers in parentheses are the associated standard deviations. ``IM'' means insufficient memory for the execution.}
	\footnotesize
	\begin{tabular}{c|cccccc}
		\hline
		Total number  & \multicolumn{6}{c}{Number of MPI processes per node} \\
		of MPI processes & 1     & 2     & 4     & 8     & 16    & 32 \\\hline
		Serial     & \textcolor[rgb]{ 1,  0,  0}{IM} &       &       &       &       &  \\
		2     & \textcolor[rgb]{ 1,  .753,  0}{173.42(1.4)} & \textcolor[rgb]{ 1,  0,  0}{IM} &       &       &       &  \\
		4     & \textcolor[rgb]{ 0,  .69,  .941}{98.04(1.04)} & \textcolor[rgb]{ 1,  .753,  0}{84.14(2.54)} & \textcolor[rgb]{ 1,  0,  0}{IM} &       &       &  \\
		8     & \textcolor[rgb]{ .329,  .51,  .208}{63.15(1.19)} & \textcolor[rgb]{ 0,  .69,  .941}{56.63(1.61)} & \textcolor[rgb]{ 1,  .753,  0}{65.44(1.63)} & \textcolor[rgb]{ 1,  0,  0}{IM} &       &  \\
		16    & \textcolor[rgb]{ .776,  .349,  .067}{36.99(0.37)} & \textcolor[rgb]{ .329,  .51,  .208}{33.00(1.12)} & \textcolor[rgb]{ 0,  .69,  .941}{35.93(3.13)} & \textcolor[rgb]{ 1,  .753,  0}{48.19(1.20)} & \textcolor[rgb]{ 1,  0,  0}{IM} &  \\
		32    & \textcolor[rgb]{ .439,  .188,  .627}{18.43(0.64)} & \textcolor[rgb]{ .776,  .349,  .067}{16.56(0.80)} & \textcolor[rgb]{ .329,  .51,  .208}{19.92(1.59)} & \textcolor[rgb]{ 0,  .69,  .941}{27.86(3.48)} & \textcolor[rgb]{ 1,  .753,  0}{37.90(1.60)} & \textcolor[rgb]{ 1,  0,  0}{IM} \\
		64    & \textcolor[rgb]{ .125,  .216,  .392}{10.34(0.34)} & \textcolor[rgb]{ .439,  .188,  .627}{10.00(0.49)} & \textcolor[rgb]{ .776,  .349,  .067}{11.92(0.27)} & \textcolor[rgb]{ .329,  .51,  .208}{15.06(1.03)} & \textcolor[rgb]{ 0,  .69,  .941}{22.38(1.11)} & \textcolor[rgb]{ 1,  .753,  0}{34.45(0.86)} \\
		\hline
	\end{tabular}
	\label{tab:AMSR-MPI-prediction}
\end{table}%

\begin{table}[htbp]
	\centering
	\caption{Mean running time (seconds) of making predictions in the parallel implementation using MPI only with different MPI process settings for the medium AMSR data set on only one Cheyenne node with 109 GB usable memory. Each experiment is run five times. Numbers in parentheses are the associated standard deviations.}
	\begin{tabular}{cccccc}
		\hline
		\multicolumn{6}{c}{Number of MPI processes} \\
		1     & 2     & 4     & 8     & 16    & 32\\
		\hline
		\textcolor[rgb]{ 1,  0,  0} {311.28(2.19)} & \textcolor[rgb]{ 1,  0,  0}{157.93(7.22)} & \textcolor[rgb]{ 1,  0,  0}{99.71(7.40)} & \textcolor[rgb]{ 1,  0,  0}{86.69(4.27)} & \textcolor[rgb]{ 1,  0,  0}{68.51(3.51)} & \textcolor[rgb]{ 1,  0,  0}{59.38(0.90)} \\
		\hline
	\end{tabular}%
	\label{tab:AMSR-MPI-prediction-large-node}%
\end{table}%

As introduced in Section~\ref{subsec:parallelMech}, each node can be utilized more efficiently if we apply shared-memory multiprocessing mechanism to each MPI process through OpenMP using the extra idle cores. The results for likelihood evaluation from the hybrid parallel implementation with MPI and OpenMP executed on the Cheyenne nodes with the 45 GB usable memory are shown in Table~\ref{tab:AMSR-hybrid-likelihood}. The strategy of the core assignments for OpenMP is given in Table~\ref{tab:coreAssignment}. Note that the settings in Table~\ref{tab:AMSR-MPI-likelihood} only constitutes half of the settings in Table~\ref{tab:AMSR-hybrid-likelihood}. For ease of comparisons between the two tables, values in colors are presented in accordance with the same setup in Table~\ref{tab:AMSR-MPI-likelihood}.
Note that the mean running time in all the settings except for the last column are smaller than the corresponding values with MPI-only execution in Table~\ref{tab:AMSR-MPI-likelihood}, which demonstrates the improvement from the OpenMP usage. The last column shows similar values to those in Table~\ref{tab:AMSR-MPI-likelihood} because only one core is assigned to each MPI process due to the physical core limit and multiprocessing does not come into play there.
When making predictions, we observe that the memory consumption goes beyond 45 GB if only one node is used. Therefore, we run the experiments for one node used on the Cheyenne node with 109 GB usable memory and all the other scenarios on the Cheyenne nodes with 45 GB usable memory. The results are shown in Table~\ref{tab:AMSR-hybrid-prediction}.

\begin{table}[t!]
	\centering
	\footnotesize
	\caption{Mean running time (seconds) of evaluating the likelihood in the hybrid parallel implementation using both MPI and OpenMP with different MPI process settings for the medium AMSR data set on the Cheyenne nodes with 45 GB usable memory. Each experiment is run five times. Numbers in parentheses are the associated standard deviations. The colors are used in accordance with the same setup in Table~\ref{tab:AMSR-MPI-likelihood} for ease of comparisons.}
	\begin{tabular}{c|cccccc}
		\hline
		Total number  & \multicolumn{6}{c}{Number of MPI processes per node} \\
		of nodes & 1     & 2     & 4     & 8     & 16    & 32 \\\hline
		1     & \textcolor[rgb]{ 1,  0,  0}{21.02(0.46)} & \textcolor[rgb]{ 1,  0,  0}{20.25(0.30)} & \textcolor[rgb]{ 1,  0,  0}{22.25(0.43)} & \textcolor[rgb]{ 1,  0,  0}{22.81(0.32)} & \textcolor[rgb]{ 1,  0,  0}{22.24(1.47)} & \textcolor[rgb]{ 1,  0,  0}{21.23(0.14)} \\
		2     & \textcolor[rgb]{ 1,  .753,  0}{11.91(0.23)} & \textcolor[rgb]{ 1,  .753,  0}{12.64(0.52)} & \textcolor[rgb]{ 1,  .753,  0}{13.63(0.16)} & \textcolor[rgb]{ 1,  .753,  0}{13.25(0.77)} & \textcolor[rgb]{ 1,  .753,  0}{13.94(1.59)} & \textcolor[rgb]{ 1,  .753,  0}{12.22(0.05)} \\
		4     & \textcolor[rgb]{ 0,  .69,  .941}{7.68(0.31)} & \textcolor[rgb]{ 0,  .69,  .941}{8.23(0.16)} & \textcolor[rgb]{ 0,  .69,  .941}{8.29(0.27)} & \textcolor[rgb]{ 0,  .69,  .941}{7.86(0.47)} & \textcolor[rgb]{ 0,  .69,  .941}{7.94(1.93)} & 7.86(0.25) \\
		8     & \textcolor[rgb]{ .329,  .51,  .208}{5.14(0.18)} & \textcolor[rgb]{ .329,  .51,  .208}{5.95(0.08)} & \textcolor[rgb]{ .329,  .51,  .208}{5.34(0.14)} & \textcolor[rgb]{ .329,  .51,  .208}{5.50(0.10)} & 6.46(1.16) & 5.93(0.28) \\
		16    & \textcolor[rgb]{ .776,  .349,  .067}{3.97(0.09)} & \textcolor[rgb]{ .776,  .349,  .067}{3.76(0.15)} & \textcolor[rgb]{ .776,  .349,  .067}{3.56(0.06)} & 3.62(0.24) & 3.89(0.19) & 3.83(0.20) \\
		32    & \textcolor[rgb]{ .439,  .188,  .627}{2.86(0.16)} & \textcolor[rgb]{ .439,  .188,  .627}{2.59(0.03)} & 2.43(0.13) & 2.76(0.32) & 2.67(0.11) & 3.23(0.27) \\
		64    & \textcolor[rgb]{ .125,  .216,  .392}{2.23(0.10)} & 1.89(0.07) & 1.93(0.23) & 1.90(0.16) & 2.09(0.12) & 2.77(0.28) \\
	\end{tabular}%
	\label{tab:AMSR-hybrid-likelihood}
\end{table}

\begin{table}[t!]
	\centering
	\footnotesize
	\caption{Mean running time (seconds) of making predictions in the hybrid parallel implementation using both MPI and OpenMP with different MPI process settings for the medium AMSR data set on the Cheyenne nodes with 109 GB usable memory for one node used as shown in the first row and 45 GB usable memory for the rest. Each experiment is run five times. Numbers in parentheses are the associated standard deviations. The colors are used in accordance with the same setup in Table~\ref{tab:AMSR-MPI-prediction} and Table~\ref{tab:AMSR-MPI-prediction-large-node} for ease of comparisons.}
	\begin{tabular}{c|cccccc}
		\hline
		Total number  & \multicolumn{6}{c}{Number of MPI processes per node} \\
		of nodes & 1     & 2     & 4     & 8     & 16    & 32 \\\hline
		1     & \textcolor[rgb]{ 1,  0,  0}{57.10(0.40)} & \multicolumn{1}{c}{\textcolor[rgb]{ 1,  0,  0}{57.01(0.78)}} & \multicolumn{1}{c}{\textcolor[rgb]{ 1,  0,  0}{59.92(1.16)}} & \multicolumn{1}{c}{\textcolor[rgb]{ 1,  0,  0}{61.61(1.95)}} & \multicolumn{1}{c}{\textcolor[rgb]{ 1,  0,  0}{57.77(0.26)}} & \multicolumn{1}{c}{\textcolor[rgb]{ 1,  0,  0}{57.74(0.08)}} \\
		2     & \textcolor[rgb]{ 1,  .753,  0}{33.99(0.49)} & \multicolumn{1}{c}{\textcolor[rgb]{ 1,  .753,  0}{34.47(0.87)}} & \multicolumn{1}{c}{\textcolor[rgb]{ 1,  .753,  0}{37.00(0.29)}} & \multicolumn{1}{c}{\textcolor[rgb]{ 1,  .753,  0}{35.28(0.49)}} & \multicolumn{1}{c}{\textcolor[rgb]{ 1,  .753,  0}{33.35(0.80)}} & \multicolumn{1}{c}{\textcolor[rgb]{ 1,  .753,  0}{34.28(0.47)}} \\
		4     & \textcolor[rgb]{ 0,  .69,  .941}{21.35(0.54)} & \multicolumn{1}{c}{\textcolor[rgb]{ 0,  .69,  .941}{22.10(0.47)}} & \multicolumn{1}{c}{\textcolor[rgb]{ 0,  .69,  .941}{22.80(1.04)}} & \multicolumn{1}{c}{\textcolor[rgb]{ 0,  .69,  .941}{21.43(1.11)}} & \multicolumn{1}{c}{\textcolor[rgb]{ 0,  .69,  .941}{21.20(2.24)}} & \multicolumn{1}{c}{20.77(0.68)} \\
		8     & \textcolor[rgb]{ .329,  .51,  .208}{13.56(0.28)} & \multicolumn{1}{c}{\textcolor[rgb]{ .329,  .51,  .208}{15.78(0.34)}} & \multicolumn{1}{c}{\textcolor[rgb]{ .329,  .51,  .208}{14.67(0.09)}} & \multicolumn{1}{c}{\textcolor[rgb]{ .329,  .51,  .208}{15.55(0.38)}} & \multicolumn{1}{c}{17.06(2.74)} & \multicolumn{1}{c}{15.16(0.27)} \\
		16    & \textcolor[rgb]{ .776,  .349,  .067}{10.32(0.24)} & \multicolumn{1}{c}{\textcolor[rgb]{ .776,  .349,  .067}{9.86(0.49)}} & \multicolumn{1}{c}{\textcolor[rgb]{ .776,  .349,  .067}{9.77(0.25)}} & 9.63(0.50) & 9.59(0.34) & \multicolumn{1}{c}{9.64(0.40)} \\
		32    & \textcolor[rgb]{ .439,  .188,  .627}{7.26(0.30)} & \multicolumn{1}{c}{\textcolor[rgb]{ .439,  .188,  .627}{6.41(0.13)}} & 6.64(0.24) & 6.89(0.25) & 6.93(0.34) & 7.33(0.25) \\
		64    & \textcolor[rgb]{ .125,  .216,  .392}{5.22(0.55)} & 4.46(0.27) & 4.62(0.37) & 5.11(0.36) & 4.99(0.21) & 7.56(1.24) \\
	\end{tabular}%
	\label{tab:AMSR-hybrid-prediction}
\end{table}

In addition, as stated in Section~\ref{subsec:parallelMech}, the dynamic scheduling strategy is intended to implement a more balanced workload for each worker. The results from the hybrid implementation with dynamic scheduling in both MPI and OpenMP are shown in Table~\ref{tab:AMSR-hybrid-dynamic-likelihood} for evaluating the likelihood and Table~\ref{tab:AMSR-hybrid-dynamic-prediction} for making predictions. We see that the running time in the majority of the cases from the dynamic scheduling scheme is shorter than the static scheduling scheme comparing Table~\ref{tab:AMSR-hybrid-dynamic-likelihood} with Table~\ref{tab:AMSR-hybrid-likelihood} and Table~\ref{tab:AMSR-hybrid-dynamic-prediction} with Table~\ref{tab:AMSR-hybrid-prediction}. The optimal number of MPI processes per node in terms of execution time depends on the number of nodes used. However, the results suggest that it would generally be better to use a smaller number of MPI processes per node in the dynamic scheduling scheme. The reason is that when we assign the workload to each MPI process dynamically, we use the sum of the number of observations squared in each region as the metric, which may not be optimal. However, the assignment for OpenMP thread is determined in the runtime, leading to a more balanced workload division.

\begin{table}[t!]
	\centering
	\footnotesize
	\caption{Mean running time (seconds) of evaluating the likelihood in the hybrid parallel implementation using both MPI and OpenMP with different MPI process settings through dynamic scheduling scheme for the medium AMSR data set on the Cheyenne nodes with 45 GB usable memory. Each experiment is run five times. Numbers in parentheses are the associated standard deviations.}
	\begin{tabular}{c|cccccc}
		\hline
		Total number  & \multicolumn{6}{c}{Number of MPI processes per node} \\
		of nodes & 1     & 2     & 4     & 8     & 16    & 32 \\\hline
		1     & 12.59(0.26) & 12.40(0.12) & 12.54(0.37) & 13.63(0.59) & 15.19(1.24) & 16.17(0.14) \\
		2     & 8.18(0.32) & 6.91(0.24) & 7.01(0.15) & 8.30(0.62) & 8.66(0.92) & 10.17(0.24) \\
		4     & 4.48(0.12) & 4.08(0.07) & 4.26(0.12) & 4.91(0.30) & 5.53(0.52) & 6.08(0.15) \\
		8     & 3.06(0.25) & 2.69(0.07) & 2.83(0.13) & 3.42(0.18) & 4.11(0.36) & 4.44(0.15) \\
		16    & 2.15(0.18) & 2.02(0.12) & 2.15(0.14) & 2.53(0.05) & 2.84(0.08) & 3.97(0.30) \\
		32    & 1.68(0.09) & 1.66(0.14) & 1.93(0.05) & 2.09(0.18) & 2.61(0.25) & 3.80(0.16) \\
		64    & 1.50(0.11) & 1.64(0.12) & 1.75(0.11) & 2.03(0.16) & 2.80(0.25) & 4.51(0.35) \\
	\end{tabular}%
	\label{tab:AMSR-hybrid-dynamic-likelihood}
\end{table}

\begin{table}[t!]
	\centering
	\footnotesize
	\caption{Mean running time (seconds) of making predictions in the hybrid parallel implementation using both MPI and OpenMP with different MPI process settings through dynamic scheduling scheme for the medium AMSR data set on the Cheyenne nodes with 109 GB usable memory for one node used as shown in the first row and 45 GB usable memory for the rest. Each experiment is run five times. Numbers in parentheses are the associated standard deviations.}
	\begin{tabular}{c|cccccc}
		\hline
		Total number  & \multicolumn{6}{c}{Number of MPI processes per node} \\
		of nodes & 1     & 2     & 4     & 8     & 16    & 32 \\\hline
		1     & 31.93(0.31) & 31.16(0.40) & 31.39(0.71) & 33.45(0.36) & 36.69(2.70) & 37.14(0.57) \\
		2     & 18.98(0.77) & 17.80(0.62) & 17.64(0.32) & 20.42(1.57) & 21.01(1.94) & 22.82(0.33) \\
		4     & 10.66(0.58) & 10.55(0.30) & 10.65(0.32) & 11.60(0.44) & 12.88(1.79) & 12.94(0.40) \\
		8     & 6.78(0.37) & 6.33(0.14) & 6.71(0.24) & 7.48(0.42) & 8.12(0.23) & 8.15(0.15) \\
		16    & 4.81(0.38) & 4.67(0.31) & 4.54(0.26) & 5.32(0.23) & 5.72(0.47) & 6.16(0.24) \\
		32    & 3.67(0.27) & 3.66(0.38) & 4.11(0.25) & 4.10(0.32) & 4.72(0.20) & 5.14(0.31) \\
		64    & 3.60(0.17) & 3.47(0.25) & 3.34(0.08) & 4.29(0.15) & 4.39(0.41) & 9.52(0.13) \\
	\end{tabular}%
	\label{tab:AMSR-hybrid-dynamic-prediction}
\end{table}

\subsubsection{Large MODIS Data Set}
\begin{figure}[t!]
	\centering
	\includegraphics[width=\linewidth]{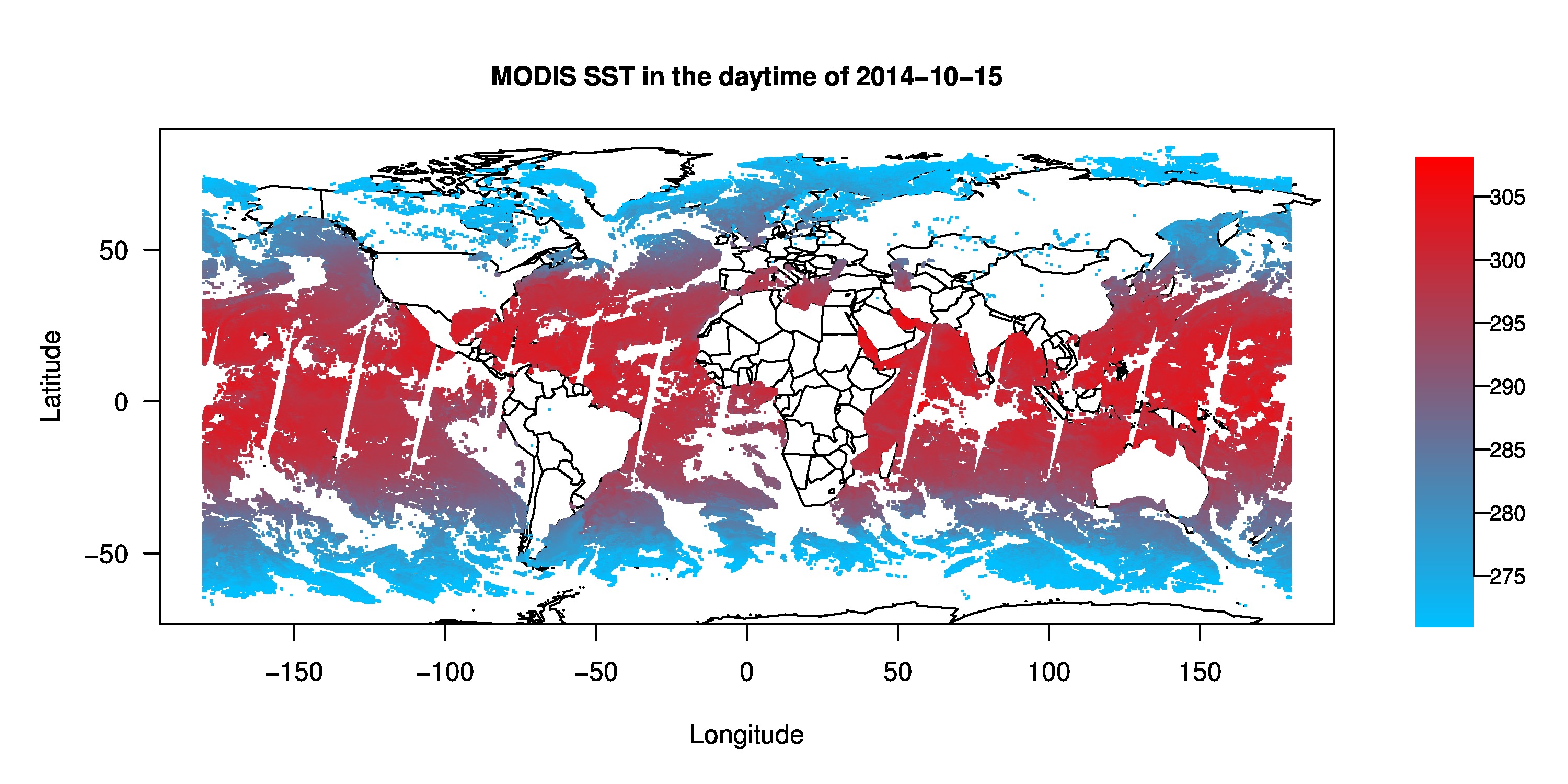}
	\caption{MODIS sea surface temperature (SST) in Kelvin in the daytime of October 15, 2014 at $47,567,759$ locations.}
	\label{fig:MODISData}
\end{figure}

There are $47,567,759$ observations for the sea surface temperature in the daytime of October 15, 2014, from
the Moderate Resolution Imaging Spectroradiometer (MODIS) sensor aboard the Aqua satellite. A visualization of the data set is given in Figure~\ref{fig:MODISData}.

\begin{figure}[ht!]
	\centering
	\includegraphics[width=\linewidth]{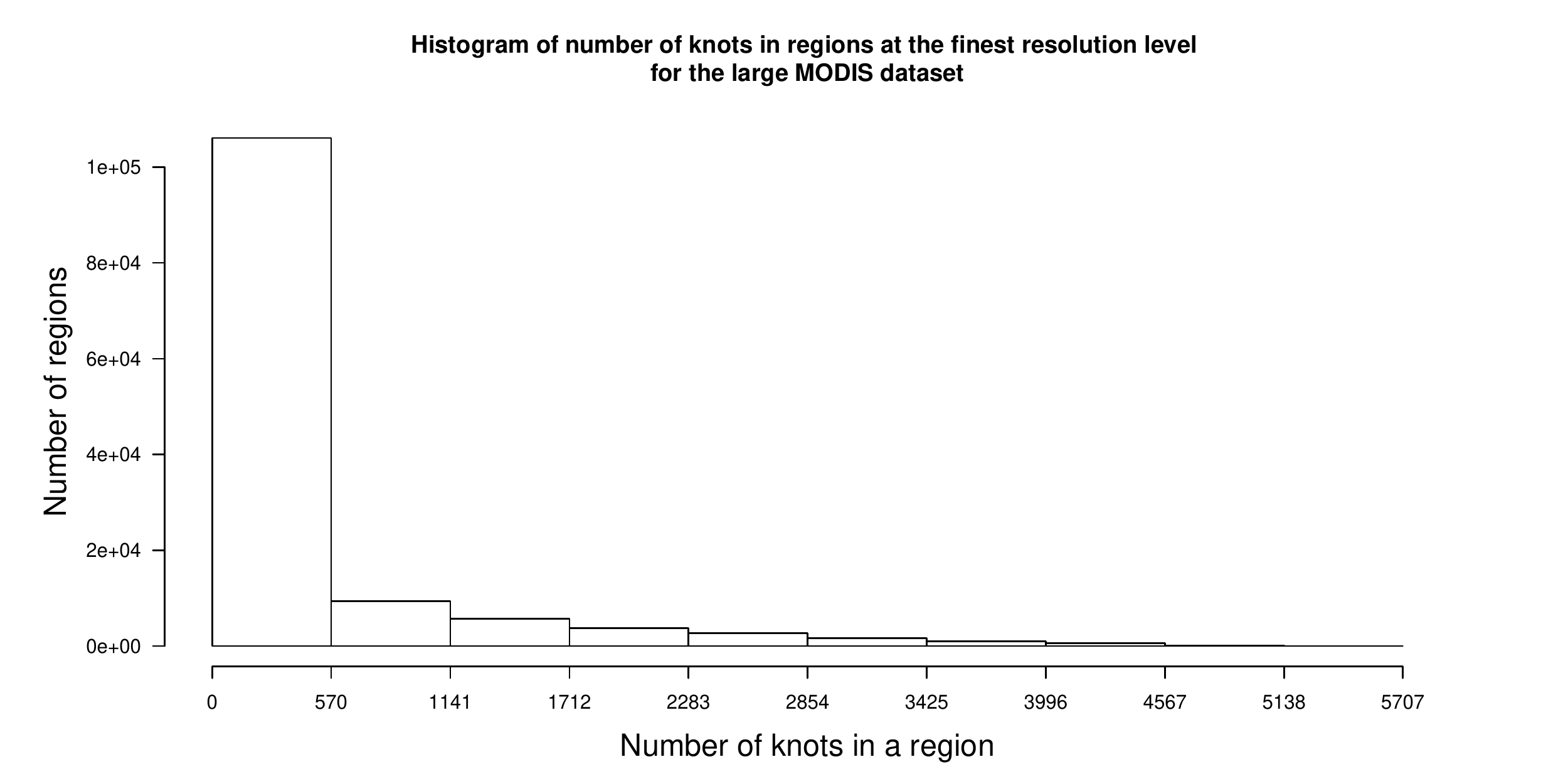}
	\caption{Histogram of the number of knots in regions at the finest resolution level for the large MODIS data set.}
	\label{fig:numberObservations-MODIS}
\end{figure}

We choose $J=2$, $M=18$ and $r=49$. Then, the largest number of observations in regions at the finest resolution level is $5,707$. The distribution of the number of observations among all the regions at the finest resolution level is skewed in a way such that a large portion of regions have smaller numbers of observations and a small portion of regions contain large numbers of observations, which is shown in Figure~\ref{fig:numberObservations-MODIS}. 

For the computations, we have shown in Section~\ref{subsubsec:amsr} that the hybrid implementation with dynamic scheduling in both MPI and OpenMP will lead to the shortest running time. So we only report the results using this strategy for this large MODIS dataset.
By monitoring the memory used, an approximate upper bound of the total memory consumption will be 720 GiB, which is too large to store on the Cheyenne nodes with 45 GB usable memory when only a relatively small number of nodes are used. We tried two approaches. The first approach is to save the most memory consuming variable \texttt{ATilde} in each region as a separate file to disk. Every time the values of \texttt{ATilde} of a particular region is needed, we load them to memory from the disk and free the allocation once finishing that region. The running time is 2663.79 seconds when using 16 nodes, each of which executes 4 MPI processes. We consider this execution is too slow to be feasible.

Another way to run the experiment is to use the Casper nodes with 384 GB memory and the Cheyenne nodes with 109 GB usable memory. Since we only have 22 Casper nodes, we can just run experiments over a few scenarios on Casper nodes. The running time results on Casper nodes in different settings are shown in Table~\ref{tab:MODIS-Casper} for evaluating likelihoods and Table~\ref{tab:MODIS-Casper-prediction} for making predictions. 
We see that for the same setting, the running time of making predictions is about 3 times the running time of evaluating likelihoods. This ratio is larger than that in Section~\ref{subsubsec:amsr}, which is 2. The reason is that the number of prediction locations here is about 47 million, considerably larger than that in Section~\ref{subsubsec:amsr}, which is about 2 million. Generally speaking, the running time is almost halved when the number of nodes used is doubled. 
The optimal number of MPI processes per node in terms of execution time is generally one or two.

The running time results on Cheyenne nodes with 109 GB usable memory for different settings are shown in Table~\ref{tab:MODIS-Cheyenne} for evaluating likelihoods and Table~\ref{tab:MODIS-Cheyenne-prediction} for making predictions with some cases failing to provide results due to insufficient memory. We see the execution on Cheyenne nodes is always faster than on Casper nodes for the same setup. The speedup factor for doubling the number of nodes used decreases as the number of used nodes increases. We have also tried to run the experiments on Cheyenne nodes with 45 GB usable memory when moderate number of nodes are used to show the execution results if computing facilities only with limited memory are available. The values in bold in Table~\ref{tab:MODIS-Cheyenne} and Table~\ref{tab:MODIS-Cheyenne-prediction} indicate that the execution can be run on Cheyenne nodes with 45 GB usable memory in the corresponding settings. It is noteworthy that in some cases indicated by the bold values running on Cheyenne nodes with 45 GB usable memory, there is a chance that the execution may crash when the memory consumption is around the limit of the memory device.

We also show the running time excluding loading the data and building the structure for likelihood evaluations in Table~\ref{tab:MODIS-single-evaluation}. In the ``optimization'' calculation mode, loading the data and building the structure would not need to be repeated. So the values of the running time for likelihood evaluations excluding these two parts indicate the execution time in each iteration of the parameter optimizations starting from the second iteration. From the results, we see it is promising that less than 10 seconds are needed for a single likelihood evaluation since the second iteration in the optimizations when using 128 Cheyenne nodes with 109 GB usable memory and two MPI processes per node for this data set on the order of 47 million observations.

\begin{table}[H]
	\centering
	\caption{Mean running time (seconds) for likelihood evaluations on Casper nodes in the hybrid parallel implementation using MPI and OpenMP with different MPI process settings through dynamic scheduling scheme for the large MODIS data set. Each experiment is run five times. Numbers in parentheses are the associated standard deviations.}
	\begin{tabular}{c|cccc}
		\hline
		Total  number& \multicolumn{4}{c}{Number of MPI processes per node} \\
		of nodes & 1 & 2 & 4 & 8 \\\hline
		4     & 410.80(11.10) & 414.33(9.66) & 422.98(10.35) & 470.23(6.75) \\
		8     & 217.55(0.69) & 218.91(1.92) & 222.49(2.39) & 248.82(4.43) \\
		16    & 118.95(0.72) & 119.85(0.30) & 122.96(1.03) & 138.98(0.60) \\\hline
	\end{tabular}%
	\label{tab:MODIS-Casper}%
\end{table}%

\begin{table}[H]
	\centering
	\caption{Mean running time (seconds) for making predictions on Casper nodes in the hybrid parallel implementation using MPI and OpenMP with different MPI process settings through dynamic scheduling scheme for the large MODIS data set. Each experiment is run five times. Numbers in parentheses are the associated standard deviations.}
	\begin{tabular}{c|cccc}
		\hline
		Total  number& \multicolumn{4}{c}{Number of MPI processes per node} \\
		of nodes & 1 & 2 & 4 & 8 \\\hline
		4     & 1,494.88(52.14) & 1,437.96(40.21) & 1,447.74(44.89) & 1,593.53(78.79) \\
		8     & 656.19(3.18) & 666.83(2.00) & 695.70(2.07) & 796.18(22.70) \\
		16    & 337.55(0.62) & 345.51(1.93) & 359.18(3.26) & 406.98(1.73) \\ \hline
	\end{tabular}%
	\label{tab:MODIS-Casper-prediction}%
\end{table}%

\begin{table}[H]
	\centering
	\caption{Mean running time (seconds) for likelihood evaluations on Cheyenne nodes with 109 GB usable memory in the hybrid parallel implementation using MPI and OpenMP with different MPI process settings through dynamic scheduling scheme for the large MODIS data set. Each experiment is run five times. Numbers in parentheses are the associated standard deviations. ``IM'' means insufficient memory for the execution. The values in bold indicate that it is also feasible to run the program for the corresponding settings on Cheyenne nodes with 45 GB usable memory.}
	\begin{tabular}{c|cccc}
		\hline
		Total  number& \multicolumn{4}{c}{Number of MPI processes per node} \\
		of nodes & 1 & 2 & 4 & 8 \\\hline
		4     & IM & IM & IM & IM \\
		8     & 139.20(0.64) & 129.05(0.52) & 133.02(1.03) & IM \\
		16    & 79.80(0.33) & 74.53(0.45) & 77.71(0.23) & 92.42(0.17) \\
		32    & 49.97(0.15) & 47.74(0.58) & 50.60(0.36) & 60.20(0.46) \\
		64    & \bf 35.08(0.15) & 36.21(0.40) & 37.86(0.21) & 45.91(0.35) \\
		128   & \bf 29.19(0.32) & \bf 29.34(0.54) & \bf 33.06(0.31) & 38.71(0.22) \\
		\hline
		
	\end{tabular}%
	\label{tab:MODIS-Cheyenne}%
\end{table}%

\begin{table}[htbp]
	\centering
	\caption{Mean running time (seconds) for making predictions on Cheyenne nodes with 109 GB usable memory in the hybrid parallel implementation using MPI and OpenMP with different MPI process settings through dynamic scheduling scheme for the large MODIS data set. Each experiment is run five times. Numbers in parentheses are the associated standard deviations. ``IM'' means insufficient memory for the execution. The value in bold indicates that it is also feasible to run the program for the corresponding setting on Cheyenne nodes with 45 GB usable memory.}
	\begin{tabular}{c|cccc}
		\hline
		Total  number& \multicolumn{4}{c}{Number of MPI processes per node} \\
		of nodes & 1 & 2 & 4 & 8 \\\hline

		4     & IM    & IM    & IM    & IM \\
		8     & IM    & IM    & IM    & IM \\
		16    & IM    & IM    & IM    & IM \\
		32    & 154.85(0.68) & 150.19(0.68) & 157.61(1.61) & IM \\
		64    & 102.73(0.64) & 102.78(0.38) & 103.89(0.45) & 130.68(0.79) \\
		128   & \bf 86.07(0.45) & 84.49(0.30) & 89.11(0.61) & 91.67(0.63) \\
		\hline

	\end{tabular}%
	\label{tab:MODIS-Cheyenne-prediction}%
\end{table}%

\begin{table}[ht]
	\centering
	\caption{Mean running time (seconds) excluding loading the data and building the structure for likelihood evaluations on Cheyenne nodes with 109 GB usable memory in the hybrid parallel implementation using MPI and OpenMP with different MPI process settings through dynamic scheduling scheme for the large MODIS data set. Each experiment is run five times. Numbers in parentheses are the associated standard deviations. ``IM'' means insufficient memory for the execution.}
	\begin{tabular}{c|cccc}
		\hline
		Total  number& \multicolumn{4}{c}{Number of MPI processes per node} \\
		of nodes & 1 & 2 & 4 & 8 \\\hline
		4     & IM & IM & IM & IM \\
		8     & 84.15(0.62) & 81.39(0.30) & 84.29(0.71) & IM \\
		16    & 44.61(0.37) & 42.62(0.37) & 44.15(0.32) & 51.42(0.50) \\
		32    & 24.43(0.17) & 23.43(0.48) & 24.13(0.33) & 29.09(0.18) \\
		64    & 13.99(0.05) & 15.08(0.35) & 14.42(0.14) & 19.04(0.04) \\
		128   & 10.02(0.15) & 9.31(0.30)  & 10.66(0.11) & 13.19(0.11) \\
		\hline
	\end{tabular}%
	\label{tab:MODIS-single-evaluation}%
\end{table}%

\newpage 

\section{Discussion}
In this article, we described our implementation of the multi-resolution approximation of Gaussian processes in \Cpp in detail, compared the performance of the serial code between \Cpp and \MATLAB for a small data set consisting of around a tenth of a million observations running on a personal laptop, studied different configurations for the parallel code applied to data sets with sizes ranging from around a tenth of a million to 47 million. Many practical concerns have been discussed to optimize the execution for the given data set and available computing facilities. In the end, we have shown that for a massive data set on the order of 47 million observations, we can get the likelihood evaluation in a single iteration of the optimization within 10 seconds (except the first iteration that needs about 30 seconds due to loading the data and building the structure) and make predictions over the 47 million observations within 85 seconds using 128 Cheyenne nodes with 109 GB usable memory.
We believe this is a great improvement that extends the use of the MRA methodology to real-world environmental applications, which often have millions of measurements.

\section*{\centering Acknowledgements}
\addcontentsline{toc}{section}{Acknowledgements}
We are grateful to Yuliya Marchetti for kindly providing the satellite data and helpful information on its usage. We would like to thank Brian Dobbins for his constructive review, in addition to having provided support and helpful advice on the use of MPI on Cheyenne. We will like to further thank Brian Vanderwende for support with \MATLAB and Matthias Katzfuss for reading an earlier version of this manuscript and providing valuable feedback.

\newpage

{
\noindent \huge \textbf{Appendices}
}

\appendix

\section{Why We Assign an Irrational Number to \texttt{offset}}\label{append:offset}
We denote $i,j$ different levels, $l_i,l_j$ the associated length of $x$-axis dimension, $x_i,x_j$ the associated position of the underlying region boundary along $x$-axis on the left, $n_i,n_j$ the associated number of knots that are placed along $x$-axis, and $f$ \texttt{offset}. Then, if two knots at levels $i$ and $j$ have the same location, it means the following equations satisfies

\[
x_i+fl_i+pl_i\dfrac{1-2f}{n_i-1} = x_j+fl_j+ql_j\dfrac{1-2f}{n_j-1}, ~\hbox{for}~ p=0,\ldots,n_i-1, q=0,\ldots,n_j-1,
\]
which is equivalent to \[
f=\dfrac{(n_i-1)(n_j-1)(x_j-x_i)+(n_i-1)ql_j-(n_j-1)pl_i}
{(n_i-1)(n_j-1)(l_i-l_j)+2(n_i-1)ql_j-2(n_j-1)pl_i}.
\]
Since all the values in the denominator and numerator are rational numbers, the equation can not hold if $f$ is irrational.

It is noteworthy that even though in theory we can avoid the collocation of any knots by assigning \texttt{offset} $f$ an irrational number, the irrational number is stored with finite decimals in computers, which are essentially rational. But with enough digits for the stored \texttt{offset} $f$, we observe that the strategy works in practice.

\section{The Upper Bound of the Memory Consumption of \texttt{ATilde}} \label{append:upperBoundATilde}
\texttt{ATilde} is defined at all regions, but the memory for regions in a particular level can be de-allocated once the program moves to the level above in the posterior inference. This means the maximum number of regions for \texttt{ATilde} to be present in memory at the same time is $J^{M-1}$ at the finest resolution level. For each region at the finest resolution level, there are $M\times (M-1)/2$ matrices associated with the conditional variance for all the ancestors and conditional covariance for any pair of ancestors. Each matrix is of dimension $r^2$. The matrix elements are 64-bit real numbers in 8 bytes. Hence, the upper bound of the memory consumption of \texttt{ATilde} is,
\[
\begin{array}{rl}
&J^{M-1}\times M\times (M-1)/2 \times r^2 \times 8 \hbox{~Bytes}\\
=&J^{M-1}\times M\times (M-1)\times r^2\times 2^{-28} \hbox{~GiB}.
\end{array}
\]


\bibliography{reference.bib}
\bibliographystyle{chicago}

\end{document}